\documentclass[lettersize,10pt,fleqn]{article}
\usepackage{graphicx}
\usepackage{subfigure}
\usepackage{morefloats}
\usepackage{color}
\usepackage{setspace}
\usepackage{floatrow}
\usepackage{bm}
\usepackage{color}
\usepackage{float}
\usepackage{amssymb}
\usepackage{geometry}%
\usepackage{amsmath}
\usepackage{CJK} 
\usepackage[colorlinks,linkcolor=blue,citecolor=blue,bookmarks,pdfstartview=FitH]{hyperref}
\usepackage{amsthm,amsmath,amssymb}
\usepackage{mathrsfs}
\usepackage{authblk}
\usepackage{cancel}  
\usepackage{multirow}
\usepackage[bottom]{footmisc}
\begin{document}

\title{\textbf{Global simulations of kinetic-magnetohydrodynamic processes with energetic electrons in tokamak plasmas}}

\author[1,2,\thanks{jbao@iphy.ac.cn}]{J. Bao}
\author[1,2,3,\thanks{Author to whom correspondence should be addressed: wzhang@iphy.ac.cn}]{W. L. Zhang}
\author[1,2,3]{D. Li}
\author[4]{Z. Lin}
\author[5]{Z. Y. Qiu}
\author[6]{W. Chen}
\author[7]{X. Zhu}
\author[8]{J. Y. Cheng}
\author[1,2]{C. Dong}
\author[1,2]{J. T. Cao}
\date{} 
\affil[1]{\small Beijing National Laboratory for Condensed Matter Physics and CAS Key Laboratory of Soft Matter Physics, Institute of Physics, Chinese Academy of Sciences, Beijing 100190, China}
\affil[2]{\small University of Chinese Academy of Sciences, Beijing 100049, China}
\affil[3]{\small Songshan Lake Materials Laboratory, Dongguan, Guangdong 523808, China}
\affil[4]{\small University of California, Irvine, CA, 92697, USA}
\affil[5]{\small Institute for Fusion Theory and Simulation and Department of Physics, Zhejiang University, Hangzhou, 310027, People’s Republic of China}
\affil[6]{\small Southwestern Institute of Physics, PO Box 432, Chengdu 610041, People’s Republic of China}
\affil[7]{\small Advanced Energy Research Center, Shenzhen University, Shenzhen 518060, China}
\affil[8]{\small Department of Physics, University of Colorado at Boulder, Boulder, Colorado 80309, USA}

\maketitle
\begin{abstract}
The energetic electrons (EEs) generated through auxiliary heating have been found to destabilize various Alfven eigenmodes (AEs) in recent experiments, which in turn lead to the EE transport and degrade the plasma energy confinement. In this work, we propose a global fluid-kinetic hybrid model for studying corresponding kinetic-magnetohydrodynamic (MHD) processes by coupling the drift-kinetic EEs to the Landau-fluid model of bulk plasmas in a non-perturbative manner. The numerical capability of Landau-fluid bulk plasmas is obtained based on a well-benchmarked eigenvalue code MAS [Multiscale Analysis of plasma Stabilities, J. Bao et al. Nucl. Fusion accepted \cite{bao2022}], and the EE responses to the electromagnetic fluctuations are analytically derived, which not only contribute to the MHD interchange drive and parallel current but also lead to the newly kinetic particle compression with the precessional drift resonance in the leading order. The hybrid model is casted into a nonlinear eigenvalue matrix equation and solved iteratively using Newton's method. By calibrating the EE precession frequency against the particle equation of motion in general geometry and applying more realistic trapped particle distribution in the poloidal plane, MAS simulations of EE-driven beta-induced Alfven eigenmodes (e-BAE) show excellent agreements with gyrokinetic particle-in-cell simulations, and the non-perturbative effects of EEs on e-BAE mode structure, growth rate and damping rate are demonstrated. With these efforts, the upgraded MAS greatly improves the computation efficiency for plasma problems related to deeply-trapped EEs,
which is superior than initial-value simulations restricted by the stringent electron Courant condition regarding to the practical application of fast linear analysis.
\end{abstract}

\newpage
\section{Introduction}

Various Alfven eigenmodes can be destabilized by energetic particles (EPs) through abundant channels of wave-particle resonances in tokamak plasmas, and the relevant kinetic-magnetohydrodynamic (MHD) processes are of great interest from aspects of experiment, theory and simulation \cite{chen2016,fasoli2007}. In particular, the AEs driven by energetic ions (EIs) are widely observed in experiments, and most linear and nonlinear properties have been studied and understood over the past four decades \cite{heidbrink2008,heidbrink2020}, while the AEs driven by energetic electrons (EEs) are much less explored before until recently, i.e., not only the single mode dynamics of EE-driven beta-induced Alfven eigenmode (e-BAE) and EE-driven toroidal Alfven eigenmode (e-TAE) are diagnosed in HL-2A \cite{chen2010, yu2018} and in EAST \cite{hu2018, chu2018,zhao2021,chu2021}, but also the experimental evidences of the nonlinear mode-mode interaction between e-TAE and geodesic acoustic mode (GAM) have been demonstrated \cite{zhu2022}. Understanding the excitation mechanism and required plasma condition of EE-driven AEs are helpful for clarifying the relevant physics in these present-day experiments, moreover, the results can be analogous to the alpha particle physics in future fusion power plant characterized by the similarly small dimensionless orbits.

Both theoretical and numerical efforts have been made to explain the EE-driven AEs in experiments. It has been shown that the bounce-averaged dynamics of EEs are responsible for the resonant interaction with MHD modes characterized by much lower frequency than EE bounce and transit frequencies ($\omega\ll\omega_{b,h}<\omega_{t,h}$) from the perspective of first-principle gyrokinetic framework \cite{zonca2007}. Recently, the bounce-kinetic EE model is also applied to the theoretical studies of linear excitation of e-BAE \cite{ma2020, ma2021} and the nonlinear excitation of zonal flow by pump e-BAE \cite{qiu2020}, which successfully reveal that the dominant destabilizing mechanism of e-BAE is attributed to the EE precessional drift resonance, as well as the important role of resonant EEs on the nonlinear mode-mode coupling. On the other hand, both the gyrokinetic particle-in-cell (PIC) simulation \cite{cheng2016,cheng2017} and kinetic-MHD hybrid simulation \cite{wang2020,wang2020b} have been performed to study the e-BAE and e-TAE with drift-kinetic description of EE dynamics, and the deeply-trapped EE response is found to have a maximal resonance island with e-BAE \cite{cheng2016} and e-TAE \cite{wang2020} propagating along the electron diamagnetic direction, while the passing EEs are responsible for the high frequency EAE \cite{wang2020}. However, regarding to the shot-to-shot analysis in experiments, the ballooning theory relies on the scale separation and is difficult to incorporate the complex geometry, and the initial value simulation using drift-kinetic EEs suffers the stringent and unnecessary numerical constraints due to the electron Courant condition with realistic electron-ion mass ratio \cite{lee2001}. For comparison, the eigenvalue approach using magnetic coordinates is more efficient for the plasma stability analysis with realistic geometry, such as LIGKA \cite{lauber2007}, NOVA-K \cite{cheng92} and CASTOR-K \cite{borba1999}, which mostly focus on EI physics rather than EE physics. Motivated by previous theories and simulations that have provided the fundamental physics insights on the linear excitation of AEs by EE preccessional drift resonance in experiments, we construct kinetic-MHD model associated with EE physics using eigenvalue approach in this work, which combines the necessary EE drive, bulk plasma damping and realistic geometry together for analyzing and optimizing experiments with fast parameter scans.

%

MAS (Multiscale Analysis for plasma Stabilities) eigenvalue code has been developed for plasma stability analysis in experimental geometry based on a five-field Landau-fluid model description of bulk plasma, which incorporates kinetic effects into MHD modes including ion finite Larmor radius (FLR), ion and electron diamagnetic drifts and Landau dampings on the same footing with fluid responses \cite{bao2022}. In this paper, we further extend MAS formulation to include EE physics described by bounce-averaged drift-kinetic equation, and a modified deeply-trapped model is proposed to solve the EE perturbed distribution and calculate corresponding kinetic particle compression (KPC) term, which is then coupled to the existing numerical capability of bulk plasma in a non-perturbative manner. After benchmarking with first-principle gyrokinetic simulations, the upgraded MAS code is useful for explaining EE-driven AEs in relevant experiments as well as providing quasi-linear spectra for integrated modeling such as machine learning \cite{dong2021}. The remainder of this paper is organized as follows. The formulation of hybrid model that couples drift-kinetic EE and Landau-fluid bulk plasma is introduced in section \ref{2}. The EE responses in various polarization and frequency limits of electromagnetic fields are shown in section \ref{3}. In section \ref{4}, the validity regime of deeply-trapped approximation is delineated, and the verification of the implemented EE module in e-BAE simulations is presented. The conclusions are given in section \ref{5}.

\section{Physics model}\label{2}

\subsection{Drift-kinetic equation for energetic electron}\label{2.1}
Considering the smallness of realistic electron mass, we apply the drift-kinetic model for EE species that induces the necessary Landau resonance while ignores the FLR effect. The linearized drift-kinetic equation in five dimensional phase space uses guiding center position $\mathbf{R}$, magnetic moment $\mu=m_ev_\perp^2/2B_0$ and parallel velocity $v_{||}$ as independent variables, which reads
\begin{flalign}\label{Vlasov}
\begin{split}
L_0\delta f_{h} + \delta L^L f_{h0} = 0
\end{split},
\end{flalign}
where $\delta f_{h}$ and $f_{h0}$ are the EE perturbed and equilibrium distributions, $L_0$ and $\delta L^L$ are the equilibrium and linear perturbed propagators given by 
\begin{flalign*}\label{}
\begin{split}
L_0= \frac{\partial }{\partial t} + \left(v_{||}\mathbf{b_0} + \mathbf{v_d}\right)\cdot\nabla
- \frac{\mu}{m_e B_0}\mathbf{B_0^*}\cdot\nabla B_0 \frac{\partial }{\partial v_{||}}
\end{split}
\end{flalign*}
and
\begin{flalign*}\label{}
\begin{split}
\delta L^L= \left(v_{||}\frac{\mathbf{\delta B}}{B_0}+\mathbf{v_E}\right)\cdot\nabla
- \left[\frac{\mu}{m_eB_0}\mathbf{\delta B}\cdot\nabla B_0 + \frac{q_e}{m_e}\left(\frac{\mathbf{B_0^*}}{B_0}\cdot\nabla\delta \phi + \frac{1}{c}\frac{\partial\delta A_{||}}{\partial t}\right)\right]\frac{\partial }{\partial v_{||}}
\end{split},
\end{flalign*}
$q_e$ and $m_e$ are electron charge and mass, $\delta\phi$ and $\delta A_{||}$ are the electrostatic potential and parallel vector potential, $\mathbf{b_0} = \mathbf{B_0}/B_0$ is the unit vector of the equilibrium magnetic field, $\mathbf{B_0^*} = \mathbf{B_0} + \left(B_0v_{||}/\Omega_{ce}\right)\nabla\times\mathbf{b_0}$, $\Omega_{ce}=q_eB_0/cm_e$ is the electron cyclotron frequency, $\mathbf{\delta B} = \nabla\times \left(\delta A_{||}\mathbf{b_0}\right)$ represents the perturbed magnetic field, and $\mathbf{v_d}$ and $\mathbf{v_E}$ are the magnetic drift and $\mathbf{E}\times \mathbf{B}$ drift, which read
\begin{flalign*}
\begin{split}
\mathbf{v_d} 
= \frac{v_{||}^2}{\Omega_{ce}}\nabla\times\mathbf{b_0} + \frac{\mu\mathbf{b_0}\times\nabla B_0}{m_e\Omega_{ce}}
\end{split}
\end{flalign*}
and 
\begin{flalign*}
\begin{split}
\mathbf{v_E} 
= \frac{c\mathbf{b_0}\times\nabla\delta\phi}{B_0}
\end{split}.
\end{flalign*}

For EE-driven Alfvenic instabilities, the linear unstable spectra can be influenced by the EE velocity distribution relying on the specific auxiliary heating methods such as ECRH and LHCD \cite{zonca2007}. Though the EE velocity distribution is not unique in experiments, we define effective EE temperatures for analysis convenience, namely, $T_{||h0} = \int m_ev_{||}^2f_{h0} \mathbf{dv}/n_{h0}$ and $T_{\perp h0} = \int \mu B_0f_{h0} \mathbf{dv}/n_{h0}$, where $n_{h0} = \int f_{h0} \mathbf{dv}$ is the EE density. In current work that illustrates the model scheme, we utilize the isotropic Maxwellian to approximate the EE equilibrium distribution, i.e., $f_{h0} = n_{h0}\left(\frac{m_e}{2\pi T_{h0}}\right)^{3/2}exp\left(-\frac{m_ev_{||}^2+2\mu B_0}{2T_{h0}}\right)$ with $T_{h0} = T_{||h0} = T_{\perp h0}$, and other EE velocity distributions such as slowing-down will be investigated in future work. We further separate the perturbed distribution $\delta f_h$ in Eq. \eqref{Vlasov} into the adiabatic part and non-adiabatic part
\begin{flalign}\label{perturbed_f}
\begin{split}
\delta f_{h} = \delta f^A + \delta K
\end{split}.
\end{flalign}

The adiabatic distribution is determined by the terms related to fast parallel dynamics in Eq. \eqref{Vlasov}, which are in the leading order of $\omega/(k_{||}v_{||})$ as
\begin{flalign}\label{adia_eq}
\begin{split}
v_{||}\mathbf{b_0}\cdot\nabla\delta f^A + v_{||}\frac{1}{B_0}\mathbf{\delta B}\cdot\nabla f_{h0}\Big{\lvert}_{v_\perp} - \frac{q_e}{m_e}\left(\mathbf{b_0}\cdot\nabla\delta\phi + \frac{1}{c}\frac{\partial \delta A_{||}}{\partial t}\right)\frac{\partial f_{h0}}{\partial v_{||}} = 0
\end{split},
\end{flalign}
where $\nabla f_{h0}{\lvert}_{v_\perp} = \left(\nabla n_{h0}/n_{h0}\right)f_{h0} + \left[\left(m_ev_{||}^2 + 2\mu B_0\right)/\left(2T_{h0}\right)-3/2\right]\left(\nabla T_{h0}/T_{h0}\right)f_{h0}$ and $\partial f_{h0}/\partial v_{||} = -\left(m_ev_{||}/T_{h0}\right)f_{h0}$ for Maxwellian $f_{h0}$. 
For analysis convenience, we newly define $\delta\psi$ in terms of $\delta A_{||}$ through relation
\begin{flalign}\label{psi}
\begin{split}
\frac{\partial \delta A_{||}}{\partial t} = -c\mathbf{b_0}\cdot\nabla\delta \psi
\end{split}.
\end{flalign}
Substituting Eq. \eqref{psi} into Eq. \eqref{adia_eq} and using the ansatz  $\partial_t = -i\omega$ and $\mathbf{b_0}\cdot\nabla = ik_{||}$, $\delta f^A$ can be readily solved from Eq. \eqref{adia_eq}
\begin{flalign}\label{adia_f}
\begin{split}
\delta f^A= - \frac{q_e}{T_{h0}}\left(\delta\phi - \delta \psi\right)f_{h0}
-\frac{q_e}{T_{h0}}\delta \psi\left[\frac{\omega_{*n,h}}{\omega}+\left(\frac{m_ev_{||}^2 + 2\mu B_0}{2T_{h0}}-\frac{3}{2}\right)\frac{\omega_{*T,h}}{\omega}\right]f_{h0}
\end{split},
\end{flalign}
where $\omega_{*n,h} = -i\frac{cT_{h0}}{q_eB_0}\mathbf{b_0}\times\frac{\nabla n_{h0}}{n_{h0}}\cdot\nabla$ and $\omega_{*T,h} = -i\frac{c}{q_eB_0}\mathbf{b_0}\times\nabla T_{h0}\cdot\nabla$. Note that the form of $\delta f^A$ in Eq. \eqref{adia_f} is also adopted in gyrokinetic theory \cite{chen1991,zonca1996} and simulation \cite{bao17}, which contains both the adiabatic response to the parallel electric field and convective response to the perturbed magnetic field.

From Eqs. \eqref{Vlasov}, \eqref{perturbed_f} and \eqref{adia_f}, the governing equation for the non-adiabatic distribution $\delta K$ can be written as
\begin{flalign}\label{non_adia_eq}
\begin{split}
&\left[-i\left(\omega - \omega_{d}\right) 
+ v_{||}\mathbf{b_0}\cdot\nabla
- \frac{\mu}{m_e B_0}\mathbf{B_0^*}\cdot\nabla B_0 \frac{\partial }{\partial v_{||}}\right]\delta K\\
&= - i\frac{q_e}{T_{h0}}\omega \left(1-\frac{\omega_{*p,h}}{\omega}\right)\left(\delta\phi - \delta\psi\right)f_{h0}
-i\frac{q_e}{T_{h0}}\omega_{d}\left(1-\frac{\omega_{*p,h}}{\omega}\right)\delta\psi f_{h0}
\end{split},
\end{flalign}
where $\omega_{d} = -i\mathbf{v_d}\cdot\nabla$ and $\omega_{*p,h}=\omega_{*n,h} + \left[\left(m_ev_{||}^2 + 2\mu B_0\right)/\left(2T_{h0}\right)-3/2\right]\omega_{*T,h}$. In nowadays tokamak plasmas, we have $C_s:V_A:v_{the} = \sqrt{\frac{\beta_e}{2}}:1:\sqrt{\frac{\beta_em_i}{2m_e}}$, where $C_s$, $V_A$ and $v_{the}$ denote ion sound speed, Alfven speed and electron thermal speed, respectively, $\beta_e = 8\pi n_eT_e/B_0^2$ and $m_e/m_i\ll\beta_e\ll 1$. With considering these orderings, we can solve $\delta K$ from Eq. \eqref{non_adia_eq} for passing and trapped EE particles, respectively. 

First, in the regime of $k_{||}v_{||}\gg\omega\sim\omega_{d}$ for passing EE particles, Eq. \eqref{non_adia_eq} reduces to
\begin{flalign}\label{non_adia_eq2}
\begin{split}
v_{||}\mathbf{b_0}\cdot\nabla\delta K^{p}
=& - i\frac{q_e}{T_{h0}}\omega \left(1-\frac{\omega_{*p,h}}{\omega}\right)\left(\delta\phi - \delta\psi\right)f_{h0}
-i\frac{q_e}{T_{h0}}\omega_{d}\left(1-\frac{\omega_{*p,h}}{\omega}\right)\delta\psi f_{h0}
\end{split},
\end{flalign}
where we have $\delta K^{p}\sim 0$ since $v_{||}\mathbf{b_0}\cdot\nabla\to\infty$, and $\delta K^{p}$ contribution to perturbed density and pressure can be ignored. However, $v_{||}\delta K^{p}$ is finite and gives rise to a fraction of parallel current density carried by passing EEs.

Second, EE particles interact and exchange energy with MHD modes primarily through the precessional drift resonance, while the bounce and transit frequencies are too high to resonate with typical AEs, as demonstrated and confirmed by recent theories and simulations \cite{ma2020, qiu2020,cheng2016, wang2020}. According to these characteristics, we have $\omega\sim\overline{\omega}_{d}\ll\omega_{b}$ for trapped EE particles, where $\omega_{b} = 2\pi/\tau_{b}$ and $\tau_{b} = \oint \left(dl/v_{||}\right)$ is the bounce period, and $\overline{\omega}_{d} = \oint \omega_{d}\left(dl/v_{||}\right)/\tau_{b}$ is the precession frequency, and $l$ is the traveled distance of trapped particle along magnetic field line in one bounce motion period. Defining $ \omega_{b}\partial/\partial\eta = v_{||}\mathbf{b_0}\cdot\nabla$ for trapped particle with $\eta$ being the bounce angle coordinate and $d\eta = \omega_{b}\left(dl/v_{||}\right)$, we can transform Eq. \eqref{non_adia_eq} into banana orbit center frame using $\delta K^t = \delta K^t_b exp\left(i\alpha\right)$ with $\alpha = -\int d\eta\left(\omega_{d}-\overline{\omega}_{d}\right)/\omega_{b}$ \cite{cai1993}, and then perform bounce-average $\overline{\left(\cdots\right)}= \oint \left(\cdots\right)\left(dl/v_{||}\right)/\tau_{b}$, i.e., $\overline{\left(\cdots\right)} = \oint \left(\cdots\right)d\eta/2\pi$. Note that  $\omega_{d}-\overline{\omega}_{d}\ll\omega_{b}$, we have $exp\left(i\alpha\right)\approx1$ which indicates that $\delta K^t \simeq \delta K^t_b$ and finite orbit width (FOW) can be dropped for EE particles as a higher order effect, then the solution of trapped EE non-adiabatic response is

\begin{flalign}\label{deltaK}
\begin{split}
\delta K^t \simeq \delta K^t_b
=
\underbrace{ \frac{\omega}{\omega-\overline{\omega}_{d}}\frac{q_e}{T_{h0}}\left(1-\frac{\omega_{*p,h}}{\omega}\right)\left(\overline{\delta\phi} -\overline{\delta\psi}\right)f_{h0}}_{\{I\}}
\underbrace{+ \frac{1}{\omega-\overline{\omega}_{d}}\frac{q_e}{T_{h0}}\left(1-\frac{\omega_{*p,h}}{\omega}\right)\overline{\omega_{d}\delta\psi} f_{h0} }_{\{II\}}
\end{split},
\end{flalign}
where term \{I\} corresponds to the non-ideal MHD effect $\Delta \phi = \delta\phi-\delta\psi$, and term \{II\} drives the bad-curvature instabilities with a minimum threshold $\omega_{*p,h}>\omega$. For most instabilities in tokamak characterized with 'flute-like' mode structures, $\delta\phi$ and $\delta\psi$ peak around the rational surfaces where $|nq - m|\ll 1$, then we can further simplify Eq. \eqref{deltaK} using $\overline{\delta\phi} \approx \delta\phi$, $\overline{\delta\psi} \approx \delta\psi$, and $\overline{\omega_{d}\delta\psi}\approx\overline{\omega}_{d}\delta\psi$. The more general validity regime of these simplifications is $exp\left[-i\left(m-nq\right)\theta\right]\approx 1$ \cite{qiu2020}, which only requires $|\theta|\ll 1/|nq-m|$. Thus, for instabilities with moderate $|nq-m|\leq1$ that deviate from rational surface, one can reduce trapped EE fraction by decreasing the $\theta$ domain of banana orbit mirror throat to more deeply-trapped regime, which still hold $\overline{\delta\phi} \approx \delta\phi$, $\overline{\delta\psi} \approx \delta\psi$, and $\overline{\omega_{d}\delta\psi}\approx\overline{\omega}_{d}\delta\psi$ for model accuracy. (Note that these simplifications are applied for deriving the EE moments in section \ref{2.2}). 

\subsection{Energetic electron continuity equation}\label{2.2}
 To couple drift-kinetic EEs with the existing Landau-fluid model of bulk plasmas, one needs to derive the EE continuity equation by integrating Eq. \eqref{Vlasov} in velocity space using $\langle\cdots\rangle_v = \int \mathbf{dv} = \frac{2\pi B_0}{m_e}\int dv_{||}d\mu$, i.e.,
\begin{flalign}\label{ee_continuity}
\begin{split}
\frac{\partial \delta n_{h}}{\partial t} 
&+ \mathbf{v_E}\cdot\nabla n_{h0} 
+ n_{h0}\mathbf{B_0}\cdot\nabla\left(\frac{\delta u_{||h}}{B_0}\right) 
+ 2cn_{h0}\nabla\delta\phi \cdot\frac{\mathbf{b_0}\times\boldsymbol{\kappa}}{B_0}
+ \frac{c}{q_e} \nabla\left(\delta P_{||h}^A + \delta P_{\perp h}^A\right)\cdot\frac{\mathbf{b_0}\times\boldsymbol{\kappa}}{B_0}\\
&+  \left\langle \boldsymbol{v_d}\cdot\nabla\delta K\right\rangle_v
= 0
\end{split},
\end{flalign}
where
\begin{flalign}\label{nee}
	\begin{split}
		\delta n_{h} = \int \left(\delta f^A + \delta K\right) \mathbf{dv}
	\end{split},
\end{flalign}
\begin{flalign}\label{uee}
	\begin{split}
		n_{h0}\delta u_{||h} = \int v_{||}\left(\delta f^A + \delta K\right) \mathbf{dv}
	\end{split},
\end{flalign}
\begin{flalign}\label{Pee_A}
	\begin{split}
		\delta P_{||h}^A = \int m_ev_{||}^2\delta f^A \mathbf{dv}
		= - q_en_{h0}\left(\delta\phi - \delta \psi\right)
		-q_en_{h0} \left(\frac{\omega_{*n,h}}{\omega}+\frac{\omega_{*T,h}}{\omega}\right)\delta \psi
	\end{split}
\end{flalign}
and
\begin{flalign}\label{Pee_A2}
	\begin{split}
		\delta P_{\perp h}^A = \int \mu B_0\delta f^A \mathbf{dv}
		= - q_en_{h0}\left(\delta\phi - \delta \psi\right)
		-q_en_{h0}\left(\frac{\omega_{*n,h}}{\omega}+\frac{\omega_{*T,h}}{\omega}\right)\delta \psi 
	\end{split}.
\end{flalign}
Taking the moment of Eq. \eqref{adia_f}, the adiabatic components in Eqs. \eqref{nee} and \eqref{uee} are obtained as
\begin{flalign}\label{nee_A}
	\begin{split}
		\delta n_{h}^A = \int \delta f^A \mathbf{dv}
		= -\frac{q_en_{h0}}{T_{h0}}\left(\delta\phi - \delta \psi\right) 
		- \frac{q_en_{h0}}{T_{h0}}\frac{\omega_{*n,h}}{\omega}\delta\psi
	\end{split}
\end{flalign}
and 
\begin{flalign}\label{uee_A}
	\begin{split}
		\delta u_{||h}^A = \frac{1}{n_{h0}}\int v_{||}\delta f^A \mathbf{dv} = 0
	\end{split}.
\end{flalign}
Eqs. \eqref{Pee_A} - \eqref{uee_A} are the adiabatic EE moments which will be used in the hybrid model.



Then we shall derive the non-adiabatic EE moments. Since the passing EEs move much faster than Alfven speed of which response is almost adiabatic to shear Alfven wave (SAW) and ion sound wave (ISW), $\delta K^p$ in Eq. \eqref{non_adia_eq2} does not contribute to density and pressure perturbations but can lead to a finite current in high-$v_{||}$ regime. On the other hand, the precessional drift of trapped EE can be close to or below the Alfven speed,  and $\delta K^t$ in Eq. \eqref{deltaK} are responsible for the non-adiabatic EE density and pressure. Moreover, it has been shown that the deeply-trapped EEs can effectively interact with most drift-type instabilities through precessional drift resonance \cite{ma2020,cheng2016,wang2020}, thus we apply the deeply-trapped approximation for computing the precession frequency in Eq. \eqref{deltaK}
\begin{flalign}\label{deep_omega_d}
	\begin{split}
	\overline{\omega}_d \approx \omega_{d0}
	\end{split},
\end{flalign}
where $\omega_{d0}$ denotes the normal curvature component of $\omega_{d}=-i\mathbf{v_d}\cdot\nabla$ at the outer midplane ($\theta = 0$), while the geodesic curvature component does not contribute to the bounce-averaged quantity. The specific form of $\omega_{d0}$ in flux coordinates is given by Eq. \eqref{omega_d_deep}, and the validation of Eq. \eqref{deep_omega_d} is discussed in section \ref{4}. Meanwhile, with the deeply-trapped approximation, the last term in Eq. \eqref{ee_continuity} becomes
\begin{flalign}\label{vPee_NA}
	\begin{split}
		\left\langle \boldsymbol{v_d}\cdot\nabla\delta K\right\rangle_v \simeq\left\langle \boldsymbol{v_d}\cdot\nabla\delta K^t\right\rangle_v
		= i\frac{2}{T_{h0}}\omega_{D0}\delta P_{h}^{NA}
	\end{split}
\end{flalign}
where $\omega_{D0} = \frac{T_{h0}}{m_eE}\omega_{d0}$, and $\delta P_{h}^{NA} \simeq \frac{1}{2}\int_{trap} \delta K^t E\mathbf{dv}$ represents the non-adiabatic EE pressure. We use pitch angle $\lambda = \mu B_a/\left(m_eE\right)$ and energy per unit mass $E = v^2/2$ as the two dimensions $\left(\lambda,E\right)$ of velocity space instead of $\left(v_{||}, \mu\right)$, where $B_a$ is the on-axis magnetic field strength, then the velocity space integration with trapped particles can be expanded as
\begin{flalign}\label{new_dv}
\begin{split}
\int_{trap} \mathbf{dv} = \sqrt{2}\pi\frac{B_0}{B_a}\sum_{\sigma}\int_{\lambda_{low}}^{B_a/B_0}\frac{1}{\sqrt{\left(1-\lambda\frac{B_0}{B_a}\right)}}d\lambda\int_{0}^{+\infty}\sqrt{E}dE
\end{split},
\end{flalign}
where $\sigma = sign(v_{||})$, and $B_a/B_{max}\leq\lambda_{low} \leq B_a/B_0$ is the lower cutoff of trapped particle pitch angle. For Maxwellian equilibrium distribution $f_{h0} = n_{h0}\left(\frac{m_e}{2\pi T_{h0}}\right)^{3/2}exp\left(-\frac{m_eE}{T_{h0}}\right)$, the trapped particle fraction $f_t$ at $B_0$ location can be expressed as
\begin{flalign}\label{ft}
	\begin{split}
		f_t = \frac{1}{n_{h0}}\int_{trap} f_{h0}\mathbf{dv}
		=\sqrt{1-\lambda_{low}\frac{B_0}{B_a}}
	\end{split}.
\end{flalign}
By counting all trapped particles at $B_0$ location with $\lambda_{low} = B_a/B_{max}$, it is seen that $f_t = 0$ in the high field side with $B_0 = B_{max}$ and $f_t = \sqrt{1-B_{min}/B_{max}}$ in the low field side with $B_0 = B_{min}$. Compared to former gyrokinetic theory that applies $f_t \approx \sqrt{2\epsilon}$ (where $\epsilon = r/R_0$) and ignores poloidal variation \cite{Chavdarovski2009}, Eq. \eqref{ft} has both radial and poloidal dependences and reflects the realistic trapped particle distribution in the poloidal plane. In general, Eq. \eqref{new_dv} can be simplified when the integrands do not rely on $\lambda$
\begin{flalign}\label{dv_E}
	\begin{split}
		\int_{trap}\mathbf{dv} 
		=4\sqrt{2}\pi f_t\int_{0}^{+\infty}\sqrt{E}dE
	\end{split}.
\end{flalign}
Thus, with the assumption in Eq. \eqref{deep_omega_d}, it is straightforward to integrate $\delta K^t$ in Eq. \eqref{deltaK} in velocity space using Eq. \eqref{dv_E}, and the non-adiabatic density and pressure perturbations are derived explicitly
\begin{flalign}\label{nee_NA}
\begin{split}
\delta n_{h}^{NA}
=&\int\delta K\mathbf{dv}\\
=&-f_t\frac{q_en_{h0}}{T_{h0}} \left[2\left(1-\frac{\omega_{*n,h}}{\omega}+\frac{3}{2}\frac{\omega_{*T,h}}{\omega}\right)\zeta R_1\left(\sqrt{\zeta}\right)-2\frac{\omega_{*T,h}}{\omega}\zeta R_3\left(\sqrt{\zeta}\right)\right]\left(\delta\phi - \delta\psi\right)\\
&-f_t
\frac{q_en_{h0}}{T_{h0}} \left[2\left(1-\frac{\omega_{*n,h}}{\omega}+\frac{3}{2}\frac{\omega_{*T,h}}{\omega}\right)R_3\left(\sqrt{\zeta}\right)-2\frac{\omega_{*T,h}}{\omega}R_5\left(\sqrt{\zeta}\right)\right]\delta\psi
\end{split}
\end{flalign}
and
\begin{flalign}\label{Pee_NA}
\begin{split}
\delta P_{h}^{NA}
=&\frac{1}{2}\int \delta K E\mathbf{dv}\\
=&-f_tq_en_{h0} \left[\left(1-\frac{\omega_{*n,h}}{\omega}+\frac{3}{2}\frac{\omega_{*T,h}}{\omega}\right)\zeta R_3\left(\sqrt{\zeta}\right)
-\frac{\omega_{*T,h}}{\omega}\zeta R_5\left(\sqrt{\zeta}\right)\right]\left(\delta\phi - \delta\psi\right) \\
&-f_tq_en_{h0}  \left[\left(1-\frac{\omega_{*n,h}}{\omega}+\frac{3}{2}\frac{\omega_{*T,h}}{\omega}\right)R_5\left(\sqrt{\zeta}\right)
-\frac{\omega_{*T,h}}{\omega}R_7\left(\sqrt{\zeta}\right)\right]\delta\psi
\end{split},
\end{flalign}
where $\zeta = \omega/\omega_{D0}$, and the response functions are given by
\begin{flalign*}\label{}
\begin{split}
&R_1 \left(\sqrt{\zeta}\right)=  1+\sqrt{\zeta}Z\left(\sqrt{\zeta}\right),\\
&R_3 \left(\sqrt{\zeta}\right)= \frac{1}{2}+\zeta+\left(\zeta\right)^{3/2}Z\left(\sqrt{\zeta}\right),\\
&R_5 \left(\sqrt{\zeta}\right)= \frac{3}{4}+\frac{1}{2}\zeta+\zeta^2+\left(\zeta\right)^{5/2}Z\left(\sqrt{\zeta}\right),\\
&R_7\left(\sqrt{\zeta}\right) = \frac{15}{8}+\frac{3}{4}\zeta+\frac{1}{2}\zeta^2+\zeta^3+\left(\zeta\right)^{7/2}Z\left(\sqrt{\zeta}\right).
\end{split}
\end{flalign*}
Substituting Eqs. \eqref{Pee_A}-\eqref{uee_A}, \eqref{vPee_NA}, \eqref{nee_NA} and \eqref{Pee_NA} into Eq. \eqref{ee_continuity}, the non-adiabatic parallel velocity $\delta u_{||h}^{NA}$ is readily solved as
 \begin{flalign}\label{uee_NA}
\begin{split}
\delta u_{||h}^{NA} = -\frac{q_e}{T_{h0}}\frac{\omega}{k_{||}}\left(1-f_t\right)\left(1-\frac{\omega_{*n,h}}{\omega}\right)\left(\delta\phi - \delta\psi\right)
-2\frac{q_e}{T_{h0}}\frac{\omega}{k_{||}}\left(\frac{\omega_{D}}{\omega}-f_t\frac{3}{4}\frac{\omega_{D0}}{\omega}\right)\left(1-\frac{\omega_{*n,h}}{\omega}-\frac{\omega_{*T,h}}{\omega}\right)\delta\psi
\end{split},
\end{flalign}
where $\omega_{D} = \frac{T_{h0}}{m_eE}\omega_{d}$. In the derivation of Eq. \eqref{uee_NA}, the resonance terms associated with $Z\left(\sqrt{\zeta}\right)$ in Eqs. \eqref{nee_NA} and \eqref{Pee_NA} exactly cancel out with each other through Eq. \eqref{ee_continuity}, which again proves that the trapped EE particles do not carry parallel current as we assumed before. On the other hand, $\delta u_{||h}^{NA}$ can also be computed by integrating Eq. \eqref{non_adia_eq2} in velocity space with passing EE fraction, which is consistent with Eq. \eqref{uee_NA} in the leading order.

\subsection{Formulation of Landau-fluid bulk plasma and drift-kinetic energetic electron hybrid model}\label{2.3}
Next, we couple the EE moments described by Eqs. \eqref{Pee_A}-\eqref{uee_A} and \eqref{nee_NA}-\eqref{uee_NA} to the Landau-fluid model of bulk plasma in Ref. \cite{bao2022}, and then yield the fluid-kinetic hybrid model as follows
\begin{flalign}\label{vor_eq}
\begin{split}
&\left[\frac{\partial}{\partial t}\left(1+ 0.75\rho_i^2\nabla_\perp^2\right)+ i\omega_{*p,i}\right]\frac{c}{V_A^2}\nabla_\perp^2\delta\phi +\mathbf{B_0}\cdot\nabla\left(\frac{1}{B_0}\nabla_\perp^2\delta A_{||}\right)
- \frac{4\pi}{c}\boldsymbol{\delta B}\cdot\nabla\left(\frac{J_{||0}}{B_0}\right)\\
&-8\pi\left(\nabla\delta P_i+\nabla\delta P_e \right)\cdot\frac{\mathbf{b_0}\times\boldsymbol{\kappa}}{B_0}
\underbrace{-8\pi\nabla\delta  P_{h}^{A}\cdot\frac{\mathbf{b_0}\times\boldsymbol{\kappa}}{B_0}}_{\{EE-IC\}}
\underbrace{- i \frac{8\pi q_e}{cT_{h0}}\omega_{D0}\delta P_{h}^{NA}}_{\{EE-KPC\}}=0
\end{split},
\end{flalign}

\begin{flalign}\label{thermal_e_ohm}
\begin{split}
\frac{\partial\delta A_{||}}{\partial t}
=-c\mathbf{b_0}\cdot\nabla\delta\phi 
-\frac{cT_{e0}}{q_en_{e0}}\mathbf{b_0}\cdot\nabla\delta n_e
- \frac{cT_{e0}}{q_en_{e0}B_0}\boldsymbol{\delta B}\cdot\nabla n_{e0}
- \frac{cm_e}{q_e}\sqrt{\frac{\pi}{2}}v_{the}|k_{||}|\delta u_{||e} 
+ \frac{c^2}{4\pi}\eta_{||}\nabla_\perp^2\delta A_{||}
\end{split},
\end{flalign}

\begin{flalign}\label{ion_pressure}
	\begin{split}
		\frac{\partial \delta P_i}{\partial t} 
		&+ \frac{c\mathbf{b_0}\times\nabla\delta\phi}{B_0}\cdot\nabla P_{i0} 
		+2\Gamma_{i\perp} P_{i0}c\nabla\delta\phi\cdot\frac{\mathbf{b_0}\times\boldsymbol{\kappa}}{B_0}
		+\Gamma_{i||} P_{i0}\mathbf{B_0}\cdot\nabla\left(\frac{\delta u_{||i}}{B_0}\right)
		-i\Gamma_{i\perp} \omega_{*p,i}Z_in_{i0}\rho_i^2\nabla_\perp^2\delta \phi\\
		&+2\Gamma_{i\perp} P_{i0}\frac{c}{Z_i}\nabla\delta T_i\cdot\frac{\mathbf{b_0}\times\boldsymbol{\kappa}}{B_0}
		+2\Gamma_{i\perp} T_{i0}\frac{c}{Z_i}\nabla\delta P_i\cdot\frac{\mathbf{b_0}\times\boldsymbol{\kappa}}{B_0}
		+n_{i0}\frac{2}{\sqrt{\pi}}\sqrt{2}v_{thi}|k_{||}|\delta T_i
		=0
	\end{split},
\end{flalign}

\begin{flalign}\label{uipara2}
	\begin{split}
		m_in_{i0}\frac{\partial\delta u_{||i}}{\partial t}
		=&\frac{Z_in_{i0}}{q_en_{e0}}\left(\mathbf{b_0}\cdot\nabla\delta P_e + \frac{1}{B_0}\boldsymbol{\delta B}\cdot\nabla P_{e0}\right)
		- \left(\mathbf{b_0}\cdot\nabla\delta P_i + \frac{1}{B_0} \boldsymbol{\delta B}\cdot\nabla P_{i0}\right)\\
		&+ Z_in_{i0}\frac{m_e}{q_e}\sqrt{\frac{\pi}{2}}v_{the}|k_{||}|\delta u_{||e}
		- Z_in_{i0}\frac{c}{4\pi}\eta_{||}\nabla_\perp^2\delta A_{||}
	\end{split},
\end{flalign}

\begin{flalign}\label{ion_density}
	\begin{split}
		\frac{\partial \delta n_i}{\partial t} 
		&+  \frac{c\mathbf{b_0}\times\nabla\delta\phi}{B_0}\cdot\nabla n_{i0} 
		+ 2cn_{i0}\nabla\delta\phi \cdot\frac{\mathbf{b_0}\times\boldsymbol{\kappa}}{B_0}
		+ n_{i0}\mathbf{B_0}\cdot\nabla\left(\frac{\delta u_{||i}}{B_0}\right)
		- i \omega_{*p,i}\frac{Z_in_{i0}}{T_{i0}}\rho_i^2\nabla_\perp^2\delta\phi \\
		&+ \frac{2c}{Z_i}\nabla\delta P_i\cdot\frac{\mathbf{b_0}\times\boldsymbol{\kappa}}{B_0}=0
	\end{split},
\end{flalign}
where $\delta P_h^A = \delta P_{||h}^A = \delta P_{\perp h}^A$ in Eq. \eqref{vor_eq}, $Z_in_{i0} + q_e\left(n_{e0}+n_{h0}\right) = 0$, and the definitions of bulk plasma variables are consistent with Ref. \cite{bao2022}. The two-moment and three-moment Landau closures are applied to thermal electrons in Eq. \eqref{thermal_e_ohm} and thermal ions in Eq. \eqref{ion_pressure} respectively, which show good agreements with drift-kinetic theory on the response functions in the regime of $k_{||}v_{thi}\ll\omega\ll k_{||}v_{the}$ \cite{hammett1990}. To close above equation set, we also have the equations for $\delta P_e$, $\delta T_e$ and $\delta T_i$ as
\begin{flalign}\label{Pe}
	\begin{split}
		\delta P_e = \delta n_eT_{e0} + n_{e0}\delta T_e
	\end{split},
\end{flalign}
\begin{flalign}\label{Te}
	\begin{split}
		\mathbf{b_0}\cdot\nabla\delta T_e + \frac{1}{B_0}\boldsymbol{\delta B}\cdot\nabla T_{e0} = 0
	\end{split}
\end{flalign}
and
\begin{flalign}\label{Ti}
	\begin{split}
	   \delta T_i = \frac{1}{n_{i0}}\left(\delta P_i - \delta n_i T_{i0}\right)
	\end{split}.
\end{flalign}
Meanwhile, the thermal electron perturbed density $\delta n_e$ and parallel velocity $\delta u_{||e}$ in Eqs. \eqref{vor_eq}, \eqref{thermal_e_ohm} and Eq. \eqref{uipara2} are calculated through the quasi-neutrality condition and parallel Ampere's law as
\begin{flalign}\label{ne}
	\begin{split}
		\delta n_e = -\frac{Z_i}{q_e}\delta n_i  - \frac{c^2}{4\pi q_eV_A^2}\nabla_\perp^2\delta\phi
		\underbrace{- \left( \delta n_{h}^A + \delta n_{h}^{NA} \right)}_{\{EE-density\}}
	\end{split}
\end{flalign}
and
\begin{flalign}\label{ue}
	\begin{split}
		q_en_{e0}\delta u_{||e} = -Z_in_{i0}\delta u_{||i} 
		- \frac{c}{4\pi}\nabla_\perp^2\delta A_{||} 
		\underbrace{-q_en_{h0}\delta u_{||h}}_{\{EE-current\}}
	\end{split}.
\end{flalign}

\begin{figure}[H]
	\center
	\includegraphics[width=0.95\textwidth]{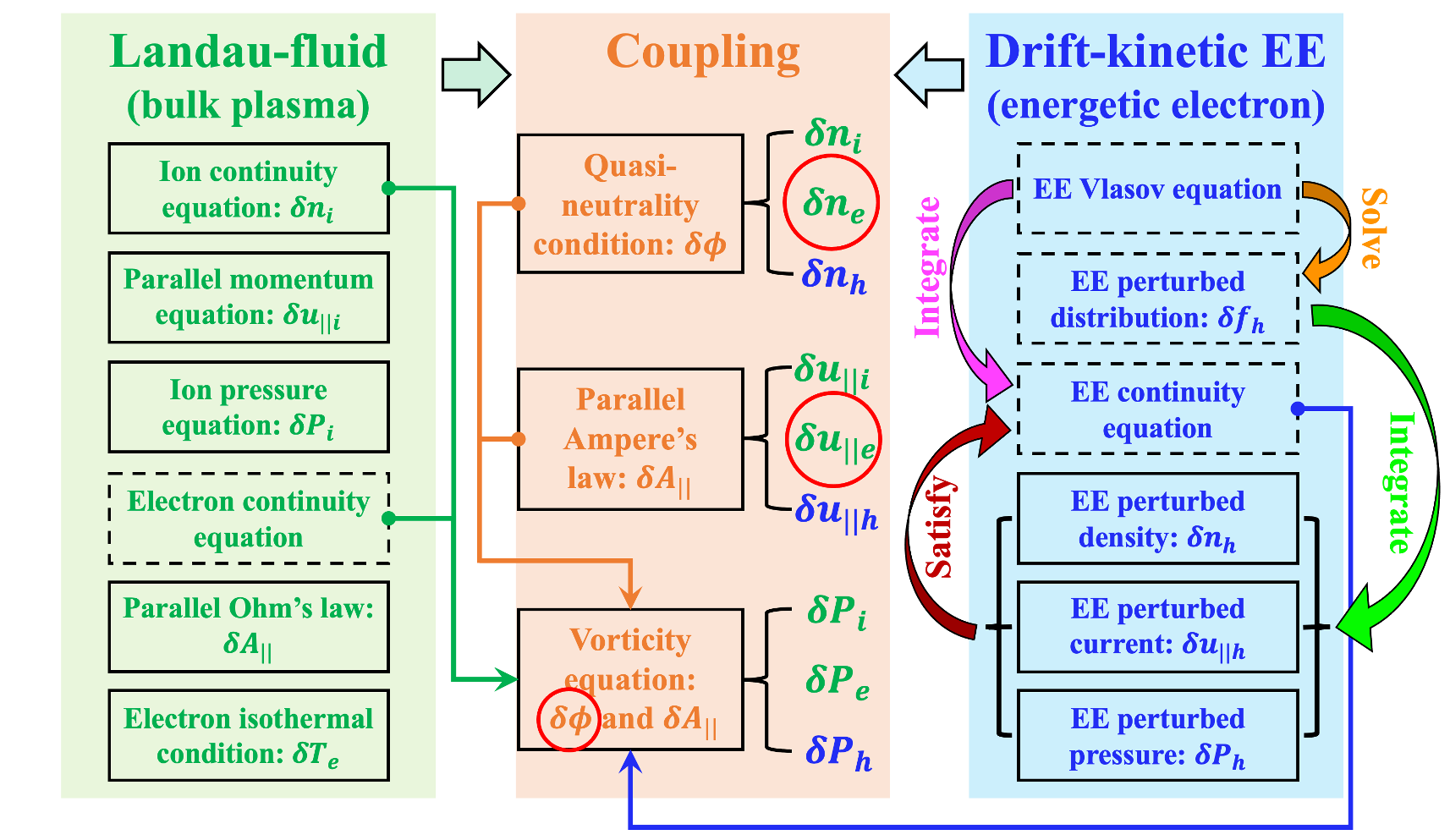}
	\caption{The schematic diagram of fluid-kinetic hybrid model. The solid boxes represent the formulation used in computation, while the equations in the dashed boxes are only used for the model derivation. The arrow lines indicate that the vorticity equation is derived by combining the thermal ion, thermal electron and EE continuity equations through the quasi-neutrality condition and parallel Ampere's law. The red circles label the unknown physical variables calculated in corresponding coupling equations. }
	\label{sketch}	
\end{figure}

Eqs. \eqref{psi}, \eqref{Pee_A}-\eqref{uee_A}, \eqref{nee_NA}-\eqref{uee_NA} and \eqref{vor_eq}-\eqref{ue} form a closed system for the fluid-kinetic hybrid simulation model as shown in figure \ref{sketch}. The main characteristics are briefly summarized here, first the well-circulating and deeply-trapped approximations on EE moments do not break the continuity equation, second the EE and bulk plasma are coupled through the quasi-neutrality condition, parallel Ampere's law and vorticity equation in a non-perturbative manner, third the EE-KPC term is responsible for the dissipative excitation of AEs while the EE-IC term contributes to the reactive MHD interchange drive in Eq. \eqref{vor_eq}.

\section{Comparison of the moment ordering between energetic electron and thermal electron}\label{3}
To estimate the ordering of each EE moment in the hybrid model, one need to compare $\delta n_{h}$, $\delta u_{||h}$ and $\delta P_{h}$ with corresponding thermal electron moments $\delta n_e$, $\delta u_{||e}$ and $\delta P_e$. Note that Eq. \eqref{ne} for $\delta n_e$ is not intuitive to compare with Eqs. \eqref{nee_A} and \eqref{nee_NA}, we solve $\delta n_e$ by using Eqs. \eqref{psi} and \eqref{thermal_e_ohm} equivalently
\begin{flalign}\label{thermal_ne}
\begin{split}
\delta n_e = -\frac{q_en_{e0}}{T_{e0}} \delta\phi
+\frac{q_en_{e0}}{T_{e0}}\left(1-\frac{\omega_{*n,e}}{\omega}\right) \delta\psi
\end{split}.
\end{flalign}
Then the thermal electron pressure $\delta P_e$ is derived using Eqs. \eqref{Te} and \eqref{thermal_ne} as
\begin{flalign}\label{thermal_pe}
\begin{split}
\delta P_e = \delta n_e T_{e0} + n_{e0}\delta T_e 
= -q_en_{e0}\delta\phi  + q_en_{e0}\left(1 - \frac{\omega_{*n,e}}{\omega} - \frac{\omega_{*T,e}}{\omega}\right)\delta\psi
\end{split},
\end{flalign}
where $\omega_{*n,e} = -i\frac{cT_{e0}}{q_en_{e0}B_0}\mathbf{b_0}\times\nabla n_{e0}\cdot\nabla$ and $\omega_{*T,e} = -i\frac{c}{q_eB_0}\mathbf{b_0}\times\nabla T_{e0}\cdot\nabla$. The thermal electron continuity equation can be obtained from Eqs. \eqref{ee_continuity}, \eqref{vor_eq}, \eqref{ion_density}, \eqref{ne} and \eqref{ue}
\begin{flalign}\label{te_density}
	\begin{split}
		\frac{\partial \delta n_e}{\partial t} 
		&+  \frac{c\mathbf{b_0}\times\nabla\delta\phi}{B_0}\cdot\nabla n_{e0} 
		+ 2cn_{e0}\nabla\delta\phi \cdot\frac{\mathbf{b_0}\times\boldsymbol{\kappa}}{B_0}
		+ n_{e0}\mathbf{B_0}\cdot\nabla\left(\frac{\delta u_{||e}}{B_0}\right)
		+ \frac{2c}{q_e}\nabla\delta P_e\cdot\frac{\mathbf{b_0}\times\boldsymbol{\kappa}}{B_0}=0
	\end{split}.
\end{flalign}
Substituting Eqs. \eqref{thermal_ne} and \eqref{thermal_pe} into Eq. \eqref{te_density} and considering $\partial_t=-i\omega$ and $\nabla = i\mathbf{k}$, we have
\begin{flalign}\label{thermal_ue}
\begin{split}
\delta u_{||e} = -\frac{q_e}{T_{e0}}\frac{\omega}{k_{||}}\left(1-\frac{\omega_{*n,e}}{\omega}\right)\left(\delta\phi - \delta\psi\right)
-\frac{2q_e}{T_{e0}}\frac{\omega}{k_{||}}\frac{\omega_{D,e}}{\omega}\left(1-\frac{\omega_{*n,e}}{\omega}-\frac{\omega_{*T,e}}{\omega}\right)\delta\psi
\end{split},
\end{flalign}
where $\omega_{D,e} = -i\frac{cT_{e0}}{q_eB_0}\mathbf{b_0}\times\boldsymbol{\kappa}\cdot\nabla$.

\begin{figure}[H]
	\center
	\includegraphics[width=1.0\textwidth]{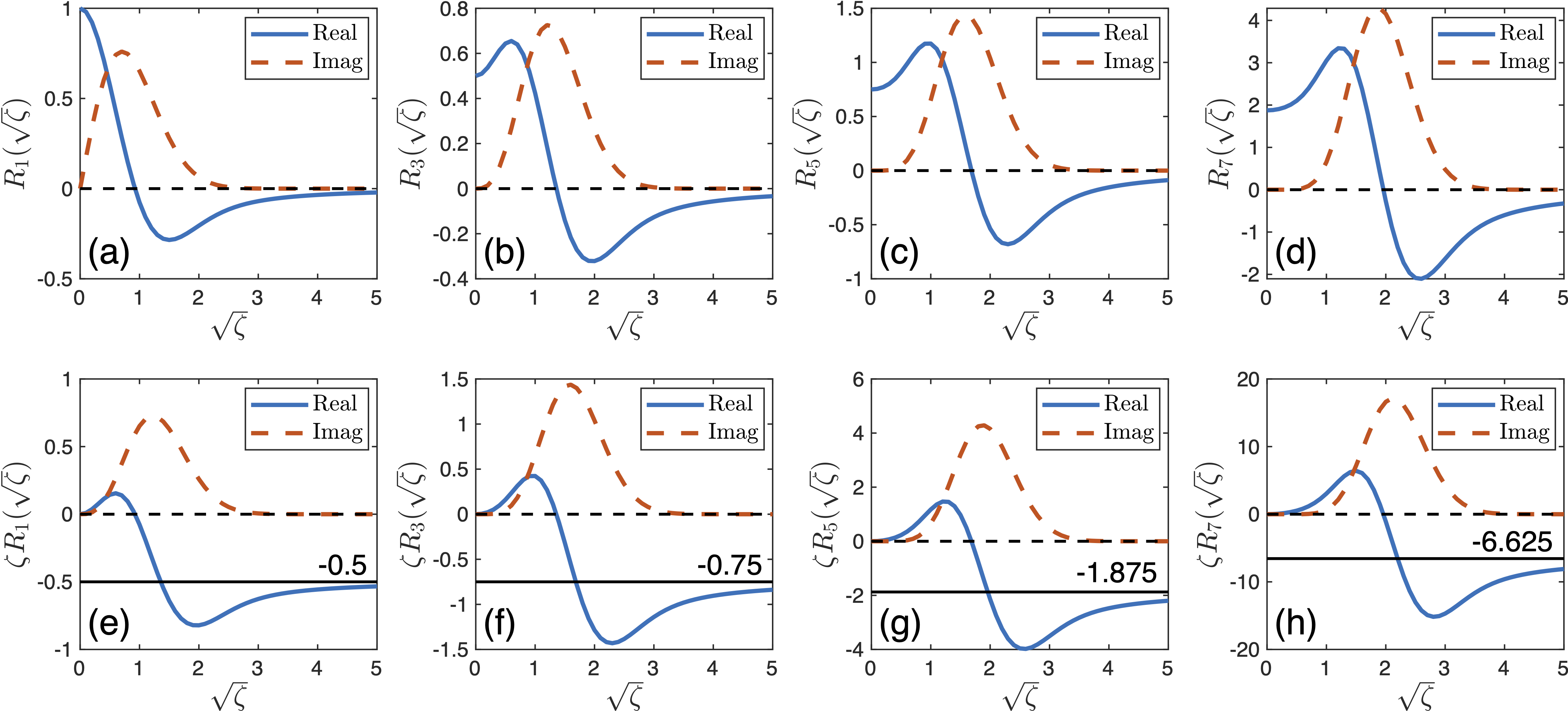}
	\caption{The real and imaginary parts of plasma response functions $R_n\left(\sqrt{\zeta}\right)$ and $\zeta R_n\left(\sqrt{\zeta}\right)$ in Eqs. \eqref{nee_NA} and \eqref{Pee_NA}. The solid black lines with the numbers in (e), (f), (g), (h) indicate the limit values at $\zeta\rightarrow+\infty $. }
	\label{R_function}	
\end{figure}

It is seen that $\delta n_{h}^A$, $\delta P_{h}^{A}$ and $\delta u_{||h}^{NA}$ in Eqs. \eqref{Pee_A}-\eqref{nee_A} and \eqref{uee_NA} already have similar forms with Eqs. \eqref{thermal_ne}, \eqref{thermal_pe} and \eqref{thermal_ue} for comparison, while $\delta n_{h}^{NA}$ and $\delta P_{h}^{NA}$ in Eqs. \eqref{nee_NA} and \eqref{Pee_NA} contain the response functions $R_n\left(\sqrt{\zeta}\right)$ and $\zeta R_n\left(\sqrt{\zeta}\right)$ as shown in figure \ref{R_function}, and we use the limit values of $\delta n_{h}^{NA}$ and $\delta P_{h}^{NA}$ at $\zeta\rightarrow 0$ and $\zeta\rightarrow+\infty$ to compare with thermal electrons, which read
\begin{flalign}\label{nee_lowzeta}
\begin{split}
\delta n_{h,0}^{NA} = \delta n_{h}^{NA}\left(\zeta \rightarrow 0\right) = -f_t\frac{q_en_{h0}}{T_{h0}}\left(1-\frac{\omega_{*n,h}}{\omega}\right)\delta\psi
\end{split},
\end{flalign}
\begin{flalign}\label{pee_lowzeta}
\begin{split}
\delta P_{h,0}^{NA} = \delta P_{h}^{NA}\left(\zeta \rightarrow 0\right) = -\frac{3}{4}f_tq_en_{h0}\left(1-\frac{\omega_{*n,h}}{\omega}-\frac{\omega_{*T,h}}{\omega}\right) \delta\psi
\end{split},
\end{flalign}

\begin{flalign}\label{nee_highzeta}
\begin{split}
\delta n_{h,+\infty}^{NA} = \delta n_{h}^{NA}\left(\zeta \rightarrow+\infty\right) = f_t\frac{q_en_{h0}}{T_{h0}}\left(1-\frac{\omega_{*n,h}}{\omega}\right)\left(\delta\phi -\delta\psi\right)
\end{split},
\end{flalign}
and
\begin{flalign}\label{pee_highzeta}
\begin{split}
\delta P_{h,+\infty}^{NA} = \delta P_{h}^{NA}\left(\zeta \rightarrow+\infty\right) = \frac{3}{4}f_tq_en_{h0}\left(1-\frac{\omega_{*n,h}}{\omega}-\frac{\omega_{*T,h}}{\omega}\right)\left(\delta\phi - \delta\psi\right)
\end{split}.
\end{flalign}

Since EE equilibrium pressure is comparable to thermal electron equilibrium pressure, we can define the smallness parameter $\delta$
\begin{flalign}\label{ordering1}
\begin{split}
\delta\sim\frac{n_{h0}}{n_{e0}}\sim\frac{T_{e0}}{T_{h0}}\sim\epsilon^2\ll 1
\end{split},
\end{flalign}
where $\epsilon = r/R_0$. Note that $L_{n,e}\sim L_{T,e} \sim L_{n,h} \sim L_{T,h}$ and $L_{n,h}/L_{B,h}\sim \epsilon$ (where $L_n = |\nabla ln \left(n\right)|^{-1}$, $L_T = |\nabla ln \left(T\right)|^{-1}$ and $L_B = |\nabla ln \left(B\right)|^{-1}$ denote the scale lengths of plasma density and temperature, and magnetic field respectively), we have
\begin{flalign}\label{ordering2}
	\begin{split}
		\frac{\omega_{D,e}}{\omega_{D}}\sim\frac{\omega_{*n,e}}{\omega_{*n,h}}\sim\frac{\omega_{*T,e}}{\omega_{*T,h}}\sim\delta
	\end{split}
\end{flalign}
and 
\begin{flalign}\label{ordering3}
	\begin{split}
		\frac{\omega_D}{\omega_{*,h}}\sim\delta^{1/2}
	\end{split},
\end{flalign}
where $\omega_{*,h} = \omega_{*n,h} + \omega_{*T,h}$. Moreover, the EE precessional drift resonance requires
\begin{flalign}\label{ordering4}
	\begin{split}
	   \omega\sim\omega_D
	\end{split}.
\end{flalign}

Then the orders of EE moments can be estimated based on Eqs. \eqref{ordering1}-\eqref{ordering4}. In the electrostatic limit $\delta\psi = 0$, we have
\begin{flalign}\label{}
\begin{split}
\frac{\delta n_h^A}{\delta n_e} 
\sim \frac{n_{h0}}{n_{e0}}\frac{T_{e0}}{T_{h0}}
\sim O\left(\delta ^2\right)
\end{split},
\end{flalign}
\begin{flalign}\label{}
\begin{split}
\frac{\delta n_{h,0}^{NA}}{\delta n_e} = 0
\end{split},
\end{flalign}
\begin{flalign}\label{}
\begin{split}
\frac{\delta n_{h,+\infty}^{NA}}{\delta n_e}
\sim -f_t\frac{n_{h0}}{n_{e0}}\frac{T_{e0}}{T_{h0}}\left(1-\frac{\omega_{*n,h}}{\omega}\right)
\sim O\left(\delta ^{3/2}\right)
\end{split},
\end{flalign}
\begin{flalign}\label{}
\begin{split}
\frac{n_{h0}\delta u_{||h}}{n_{e0}\delta u_{||e}}
\sim \left(1-f_t\right)\frac{n_{h0}T_{e0}}{n_{e0}T_{h0}}\left(\frac{\omega-\omega_{*n,h}}{\omega - \omega_{*n,e}}\right)
\sim O\left(\delta ^{3/2}\right)
\end{split},
\end{flalign}
\begin{flalign}\label{}
\begin{split}
\frac{\delta P_h^A}{\delta P_e} = \frac{n_{h0}}{n_{e0}}
\sim O\left(\delta\right)
\end{split},
\end{flalign}
\begin{flalign}\label{}
\begin{split}
\frac{\delta P_{h,0}^{NA}}{\delta P_e} = 0
\end{split}
\end{flalign}
and
\begin{flalign}\label{}
\begin{split}
\frac{\delta P_{h,+\infty}^{NA}}{\delta P_e}
\sim -\frac{3}{4}f_t\frac{n_{h0}}{n_{e0}}\left(1-\frac{\omega_{*n,h}}{\omega}-\frac{\omega_{*T,h}}{\omega}\right)
\sim O\left(\delta ^{1/2}\right)
\end{split}.
\end{flalign}
It is seen that only $\delta P_h^{NA}$ is probably close to $\delta P_e$ when $\zeta\to\infty$, however, the vorticity Eq. \eqref{vor_eq} that contains $\delta P_h^{NA}$ term becomes redundant in the electrostatic limit, thus EEs are not important to the electrostatic polarized modes. In the electromagnetic limit $\delta\phi = \delta\psi$, we have
\begin{flalign}\label{}
\begin{split}
\frac{\delta n_h^A}{\delta n_e} 
\sim \frac{n_{h0}}{n_{e0}}\frac{T_{e0}}{T_{h0}}\frac{\omega_{*n,h}}{\omega_{*n,e}}
\sim O\left(\delta\right)
\end{split},
\end{flalign}
\begin{flalign}\label{}
\begin{split}
\frac{\delta n_{h,0}^{NA}}{\delta n_e} 
\sim f_t\frac{n_{h0}}{n_{e0}}\frac{T_{e0}}{T_{h0}}\frac{\omega - \omega_{*n,h}}{\omega_{*n,e}}
\sim O\left(\delta\right)
\end{split},
\end{flalign}
\begin{flalign}\label{}
\begin{split}
\frac{\delta n_{h,+\infty}^{NA}}{\delta n_e}
= 0
\end{split},
\end{flalign}
\begin{flalign}\label{}
\begin{split}
\frac{n_{h0}\delta u_{||h}}{n_{e0}\delta u_{||e}}
\sim \frac{n_{h0}T_{e0}}{n_{e0}T_{h0}}\left(\frac{\omega_{D} - \frac{3}{4}f_t\omega_{D0}}{\omega_{D,e}}\right)\left(\frac{\omega - \omega_{*n,h}-\omega_{*T,h}}{\omega - \omega_{*n,e}-\omega_{*T,e}}\right)
\sim O\left(\delta ^{1/2}\right)
\end{split},
\end{flalign}
\begin{flalign}\label{}
\begin{split}
\frac{\delta P_{h}^{A}}{\delta P_e} 
\sim \frac{n_{h0}}{n_{e0}}\left(\frac{\omega_{*n,h} + \omega_{*T,h}}{\omega_{*n,e} + \omega_{*T,e}}\right)
\sim 1
\end{split},
\end{flalign}
\begin{flalign}\label{}
\begin{split}
\frac{\delta P_{h,0}^{NA}}{\delta P_e} = 0
\end{split}
\end{flalign}
and
\begin{flalign}\label{}
\begin{split}
\frac{\delta P_{h,+\infty}^{NA}}{\delta P_e}
\sim -\frac{3}{4}f_t\frac{n_{h0}}{n_{e0}}\left(\frac{\omega_{*n,h} + \omega_{*T,h}-\omega}{\omega_{*n,e} + \omega_{*T,e}}\right)
\sim 1
\end{split}.
\end{flalign}
Therefore, $\delta P_h^A$, $\delta P_h^{NA}$ and $\delta u_{||h}$ are the leading order terms in the electromagnetic limit, and should be kept for EE species in the hybrid model, while the EE density perturbations $\delta n_h^A$ and $\delta n_h^{NA}$ can be safely dropped as higher order small terms.

\section{Verification and validation of the fluid-kinetic hybrid model}\label{4}
The fluid-kinetic hybrid model in section \ref{2.3} can be casted into a nonlinear eigenvalue equation in $\omega$ as
\begin{flalign}\label{matrix_eq}
	\begin{split}
	     \mathbb{A}\mathbf{X} - \omega \mathbb{B}\mathbf{X} + \mathbb{C}\left(\omega\right)\mathbf{X} = 0
	\end{split},
\end{flalign}
where $\mathbb{A}$ and $\mathbb{B}$ are the operators of bulk plasma Landau-fluid model and are independent of $\omega$, and $\mathbb{C}\left(\omega\right)$ is the drift-kinetic EE operator that relies on $\omega$ nonlinearly. We use Newton's  iterative method to solve Eq. \eqref{matrix_eq} with the initial guesses of $\left(\omega, \mathbf{X}\right)$ obtained from $\mathbb{A}\mathbf{X} = \omega \mathbb{B}\mathbf{X}$, which is efficient when EE term is small and perturbative. For cases that EE term is comparable with bulk plasma terms, we solve $\mathbb{A}\mathbf{X} - \omega \mathbb{B}\mathbf{X} + \epsilon\mathbb{C}\left(\omega\right)\mathbf{X} = 0$ in the middle steps instead of directly solving Eq. \eqref{matrix_eq}, where $\epsilon$ is an adjustable parameter that gradually increases from 0 to 1, thus the proper initial guesses of $\left(\omega, \mathbf{X}\right)$ can be obtained for each $\epsilon$-value case and guarantee the robustness of Newton's iterative method in the non-perturbative regime.

To verify the EE physics model and numerical scheme, we first validate the EE precession frequency that uses deeply-trapped approximation in both analytic and experimental tokamak geometries, and then carry out the verification simulations of e-BAE to demonstrate the correctness of EE precessional drift resonance.
\subsection{How good is deeply-trapped approximation for bounce-averaged guiding-center dynamics?}\label{4.1}

As claimed by Eq. \eqref{deep_omega_d} in section \ref{2}, we utilize the deeply-trapped approximation for calculating the precession frequency of trapped EE. To delineate its regime of validity, we compare Eq. \eqref{deep_omega_d} with the exact precession frequency by performing bounce-average along the realistic banana orbit. The Boozer coordinates $\left(\psi,\theta,\zeta\right)$ is applied in MAS to describe the magnetic field, which has the contravariant form $\mathbf{B_0} = q\left(\psi\right)\nabla\psi\times\nabla\theta - \nabla\psi\times\nabla\zeta$ and the covariant form $\mathbf{B_0} = I\left(\psi\right)\nabla\theta + g\left(\psi\right)\nabla\zeta$. Then the guiding-center equation of motion in equilibrium $B$-field can be expressed in $\left(\psi,\theta,\zeta\right)$ coordinates \cite{white_book}
\begin{flalign}\label{psi_dot}
\begin{split}
\dot{\psi} = -\frac{c}{Z_\alpha}\left(\frac{m_\alpha v_{||}^2}{B_0} + \mu\right)\frac{1}{JB_0^2}\left(g\frac{\partial B_0}{\partial\theta}\right)
\end{split},
\end{flalign}
\begin{flalign}\label{theta_dot}
\begin{split}
\dot{\theta} = \frac{v_{||}}{JB_0} 
- \frac{cm_\alpha v_{||}^2}{Z_\alpha}\frac{1}{JB_0^2}\frac{\partial g}{\partial\psi}
+ \frac{c}{Z_\alpha}\left(\frac{m_\alpha v_{||}^2}{B_0}+\mu\right)\frac{1}{JB_0^2}g\frac{\partial B_0}{\partial\psi}
\end{split},
\end{flalign}
\begin{flalign}\label{}
\begin{split}
\dot{\zeta} = \frac{qv_{||}}{JB_0} 
+ \frac{cm_\alpha v_{||}^2}{Z_\alpha}\frac{1}{JB_0^2}\frac{\partial I}{\partial\psi}
-\frac{c}{Z_\alpha}\left(\frac{m_\alpha v_{||}^2}{B_0}+\mu\right)\frac{1}{JB_0^2}I\frac{\partial B_0}{\partial\psi}
\end{split}
\end{flalign}
and
\begin{flalign}\label{rhopara}
\begin{split}
\dot{\rho}_{||} = -\left(1-\rho_{||}g^{\prime}\right)\frac{c}{Z_\alpha}\left(\frac{m_\alpha v_{||}^2}{B_0} + \mu\right)\frac{1}{JB_0^2}\frac{\partial B_0}{\partial\theta}
\end{split},
\end{flalign}
where $\alpha$ denotes the particle species, $J = \left(\nabla\psi\times\nabla\theta\cdot\nabla\zeta\right)^{-1}$ is the Jacobian and $\rho_{||} = v_{||}/\Omega_{c\alpha}$. Given energy $E = 0.5v^2$ and pitch angle $\lambda = \mu B_a/\left(m_\alpha E\right)$ of a particle, parallel velocity $v_{||}$ can be written as
\begin{flalign}\label{vpara}
\begin{split}
v_{||} = \pm\sqrt{2E\left(1-\lambda\frac{B_0}{B_a}\right)}
\end{split},
\end{flalign}
and the bounce period can be calculated using Eq. \eqref{theta_dot} and \eqref{vpara} \cite{white_book}
\begin{flalign}\label{tau_b}
\begin{split}
\tau_b =  \oint d\theta/\dot{\theta}
\simeq \oint\frac{JB_0}{v_{||}}d\theta
\end{split}.
\end{flalign}
Using Eqs. \eqref{psi_dot} -\eqref{rhopara} and \eqref{tau_b} , the exact precession frequency is then derived as \cite{white_book}
\begin{flalign}\label{omega_d}
\begin{split}
\overline{\omega}_d =&n\overline{\mathbf{v_d}\cdot\nabla\left(\zeta-q\theta\right)}\\
=&\frac{n}{\tau_b}\left[-\oint \frac{1}{\dot{\theta}}\frac{c}{Z_\alpha}\left(\mu + \frac{m_\alpha v_{||}^2}{B_0}\right)\frac{\partial B_0}{\partial\psi}d\theta
+ \oint\frac{cm_\alpha v_{||}}{Z_\alpha B_0}\left(\frac{\partial I}{\partial\psi} + q\frac{\partial g}{\partial\psi}\right)d\theta
+ \oint\rho_{||}g\frac{dq}{d\psi}d\theta\right]
\end{split},
\end{flalign}
where $n$ is the toroidal mode number, and the first term on the RHS is the leading order term that depends on both the radial $\psi$ and poloidal $\theta$ coordinates, i.e., $\overline{\omega}_d = \overline{\omega}_d\left(\psi,\theta\right)$.

On the other hand, the magnetic drift frequency $\omega_d$ in Boozer coordinate can be expressed as
\begin{flalign}\label{wd}
\begin{split}
\omega_d = -i\mathbf{v_d}\cdot\nabla = 
&\underbrace{i\frac{c}{Z_\alpha}\left(\mu + \frac{m_\alpha v_{||}^2}{B_0}\right)\frac{1}{JB_0^2}\left[g\frac{\partial B_0}{\partial\theta}\frac{\partial}{\partial\psi}\right]}_{\{I\}}
\underbrace{-n\frac{c}{Z_\alpha}\left(\mu + \frac{m_\alpha v_{||}^2}{B_0}\right)\frac{\partial B_0}{\partial\psi}}_{\{II\}}\\
&\underbrace{+ n\frac{cm_\alpha v_{||}^2}{Z_\alpha}\frac{1}{JB_0^2}\left(\frac{\partial g}{\partial\psi}q + \frac{\partial I}{\partial\psi}\right)}_{\{III\}}
\end{split},
\end{flalign}
where term {\{I\}}, term {\{II\}} and term {\{III\}} represent the geodesic curvature, normal curvature and equilibrium current contributions respectively. In Eq. \eqref{wd}, we note that term {\{I\}} doesn't contribute to the precession frequency since the geodesic drift cancels over a bounce period, effect of term \{III\} is ignorable, while term {\{II\}} is equal to Eq. \eqref{omega_d} in the limit of $\tau_b\to 0$, $\theta\to 0$ and $v_{||}\to 0$ (i.e., deeply-trapped limit)
\begin{flalign}\label{omega_d_deep}
\begin{split}
\omega_{d0} = \overline{\omega}_{d,\tau_b\to 0} =- n\frac{c}{Z_\alpha}\left(\mu + \frac{m_\alpha v_{||}^2}{B_0}\right)\frac{\partial B_0}{\partial\psi}\Big\lvert_{\theta=0}
\end{split},
\end{flalign}
which is used for the deeply-trapped approximation in Eq. \eqref{deep_omega_d}, and only has $\psi$ dependence, i.e., $\omega_{d0} = \omega_{d0}\left(\psi\right)$.

\begin{figure}[H]
	\center
	\includegraphics[width=0.4\textwidth]{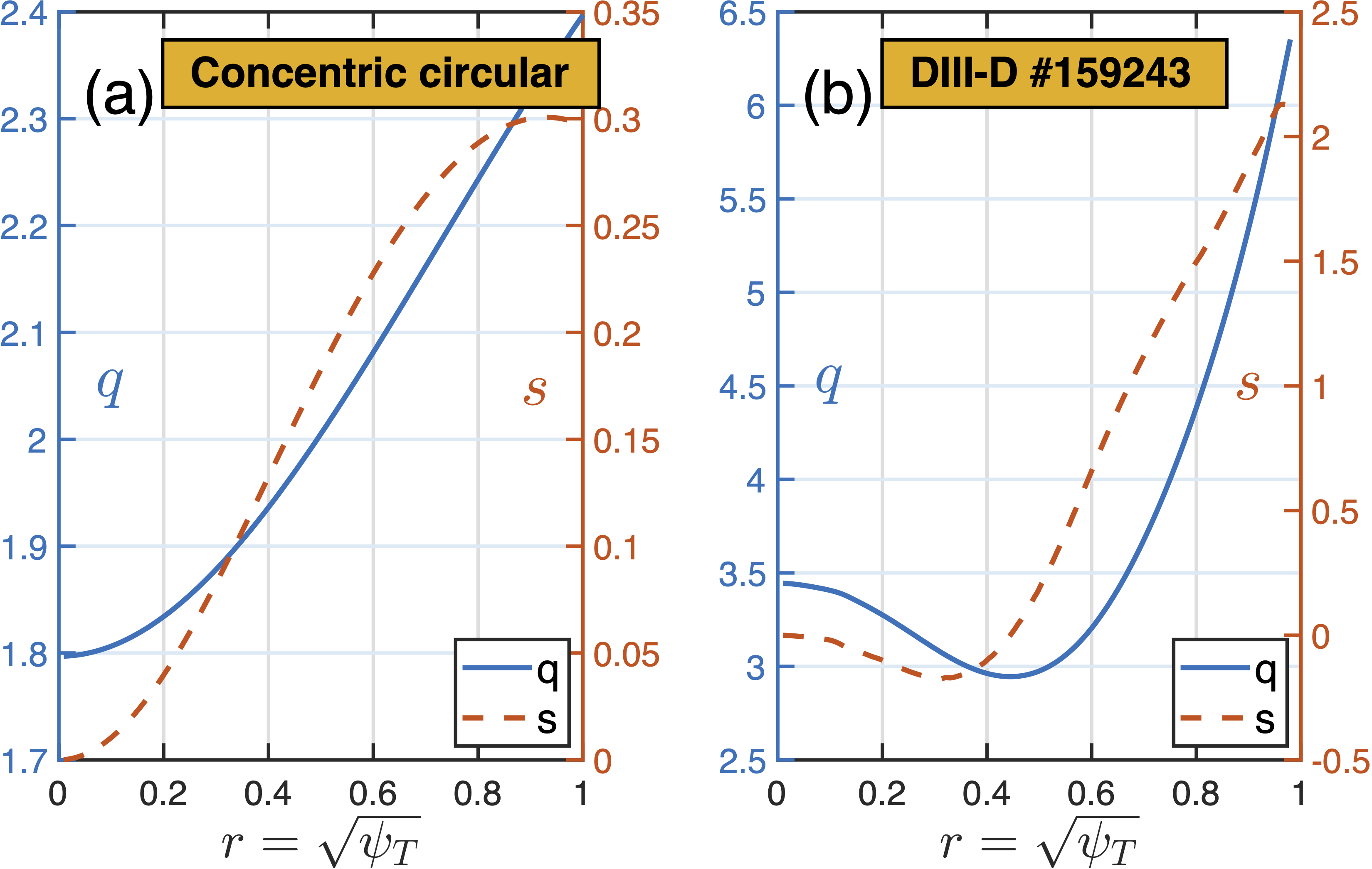}
	\caption{The safety factor $q$ profile and magnetic shear $s = \frac{1}{q}\frac{dq}{dr}$ profile in (a) analytic geometry and (b) DIII-D geometry. $\psi_T$ is the normalized toroidal magnetic flux.}
	\label{q}	
\end{figure}

\begin{figure}[H]
	\center
	\includegraphics[width=0.9\textwidth]{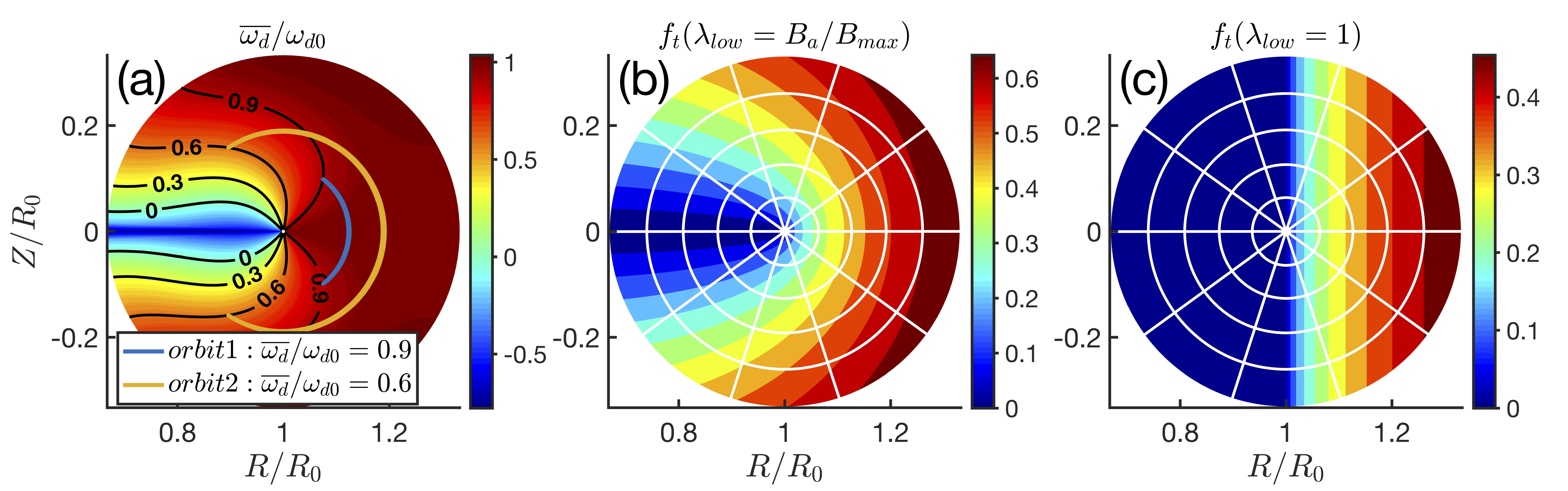}
	\caption{Analytic geometry using concentric circular. (a) The ratio between the exact precession frequency $\overline{\omega}_d\left(\psi_b,\theta_b\right)$ and the deeply-trapped approximation $\omega_{d0}\left(\psi_b\right)$ of all trapped EEs in the poloidal plane. Note that $\overline{\omega}_d/\omega_{d0}$ is plotted according to banana tip coordinates $\left(\psi_b,\theta_b\right)$. The trapped particle fraction $f_t$ in Eq. \eqref{ft} with different lower cutoffs of pitch angle (b) $\lambda_{low} = B_a/B_{max}$ and (c) $\lambda_{low} = 1$.}
	\label{ana_trap}
\end{figure}

\begin{figure}[H]
	\center
	\includegraphics[width=0.9\textwidth]{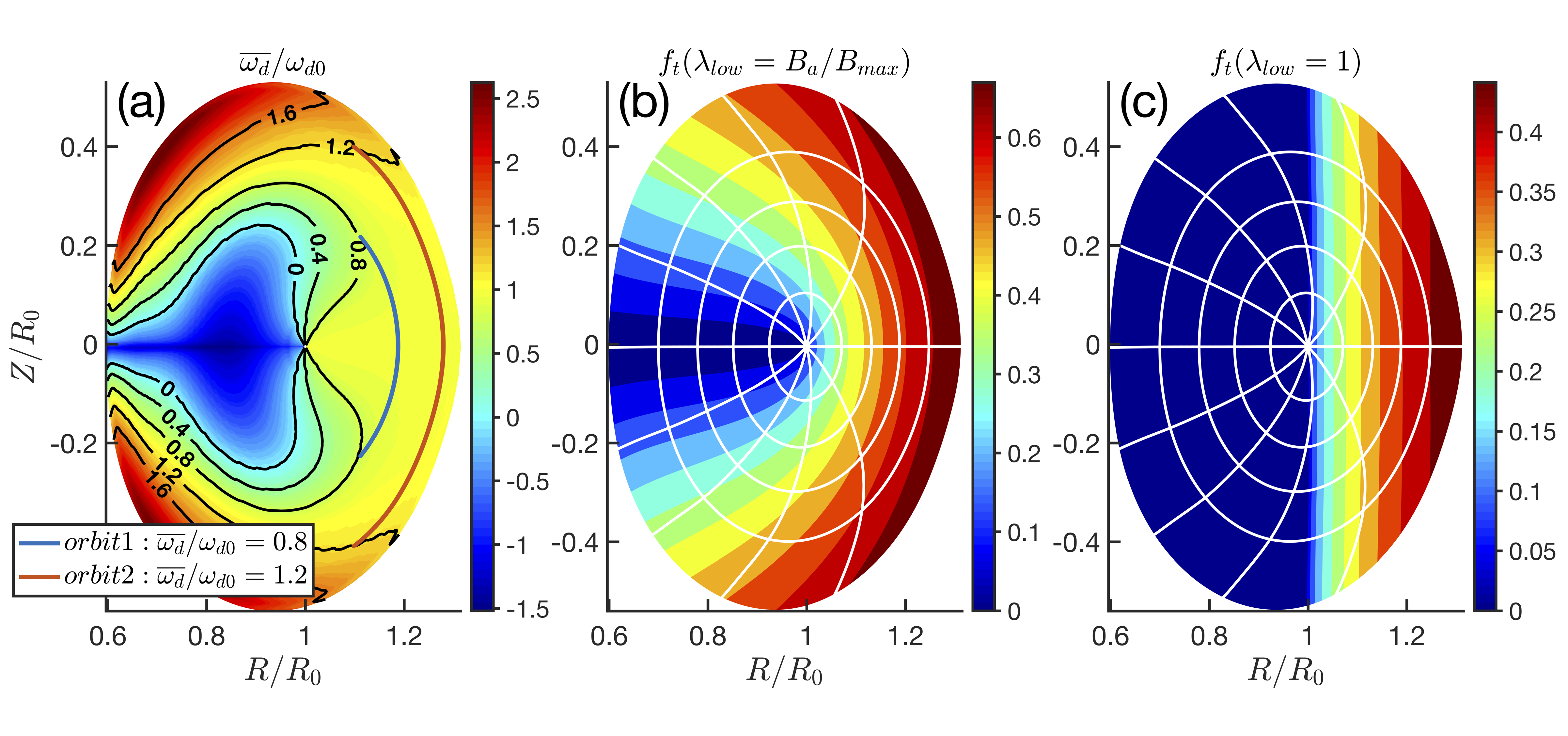}
	\caption{DIII-D general geometry. The captions of (a)-(c) are the same with figure \ref{ana_trap}.}
	\label{d3d_trap}	
\end{figure}

In order to elucidate the accuracy of $\omega_{d0}$, we compare Eqs. \eqref{omega_d} and \eqref{omega_d_deep} in both the analytic geometry \cite{cheng2016} and DIII-D general geometry \cite{sam2019}. The safety factor $q$ and magnetic shear $s = (1/q)(dq/dr)$ are given in figure \ref{q}. For analysis convenience, we use $\left(\psi_b,\theta_b\right)$ coordinates to represent trapped particles, where $\psi_b\in[0,\psi_{edge}]$ is the poloidal magnetic flux of banana orbit center and $\theta_b \in[0,2\pi]$ is the Boozer poloidal angle at the banana tip location (i.e., the mirror throat). Using the magnetic field strength at $\left(\psi_b,\theta_b\right)$ location, i.e., $B_{0,b} = B_0\left(\psi_b,\theta_b\right)$, we can express the pitch angle of trapped particle as $\lambda = B_a/B_{0,b}$, which varies in the range of $B_a/B_{max}<\lambda<B_a/B_{min}$. Substituting Eq. \eqref{vpara} into Eq. \eqref{omega_d}, the exact precession frequency $\overline{\omega}_d$ can be calculated for all trapped particles in terms of $\left(\psi_b,\theta_b\right)$. The ratio between $\overline{\omega}_d\left(\psi_b,\theta_b\right)$ and $\omega_{d0}\left(\psi_b\right)$ in the analytic and DIII-D general geometries are shown in figures \ref{ana_trap} (a) and \ref{d3d_trap} (a) respectively. First, it is seen that the numerical calculation of Eq. \eqref{omega_d} agrees with the measurement from the realistic orbits by Eqs. \eqref{psi_dot}-\eqref{rhopara}. Second, most areas on the low field sides are characterized with $\overline{\omega}_d/\omega_{d0}\sim 1$, thus $\omega_{d0}$ in Eq. \eqref{omega_d_deep} can well approximate the precession frequency for most trapped particles not only in the analytic geometry with low magnetic shear but also in the experimental geometry with broader range of magnetic shear, where the significance of magnetic shear term in Eq. \eqref{omega_d} is decreased by the elongation effect.

Moreover, since the trapped particle fraction $f_t$ determines the EE-drive intensity through Eqs. \eqref{nee_NA} and \eqref{Pee_NA}, one needs to adjust $\lambda_{low}$ in Eq. \eqref{ft} so that $f_t$ only incorporates the contribution of trapped particles that satisfy $\overline{\omega}_d/\omega_{d0}\sim 1$, while the barely-trapped particles beyond the model capability should be excluded. Besides the constraint arising from the approximation on precession frequency, we note that the bounce-average operations on electromagnetic fields in Eq. \eqref{deltaK} are removed for integrating the EE moments, and corresponding validity regime of $|\theta|\ll1/|nq-m|$ becomes another constraint for $\lambda_{low}$. Figures \ref{ana_trap} (b) and \ref{d3d_trap} (b) show the 2D profile of $f_t$ using $\lambda_{low} = B_a/B_{max}$ in the poloidal plane, which take into account all trapped particles and overestimate the deeply-trapped EE-drive intensity. We then apply $\lambda_{low} = 1$ in Eq. \eqref{ft} to calculate $f_t$ as shown in figures \ref{ana_trap} (c) and \ref{d3d_trap} (c), which corresponds to the trapped particles with entire banana orbits on the low field side, in consistency with the validity regime of $\overline{\omega}_d/\omega_{d0}\sim 1$ in figures \ref{ana_trap} (a) and \ref{d3d_trap} (a). Thus, the 2D function $f_t$ calculated by using $\lambda_{low}=1$ in Eq. \eqref{ft} correctly reflects the deeply-trapped EE-drive intensity for modes peak at $k_{||}\sim 0$ and is applied in the following e-BAE simulations. For modes peak at moderate $k_{||}\leq 1/qR_0$ such as e-TAE, one needs to choose $\lambda_{low}<1$ in Eq. \eqref{ft} to calculate $f_t$ in the more deeply-trapped regime (i.e., reduce $\theta_b$). For comparison, the widely used simple 1D form of $f_t \approx \sqrt{2\epsilon}$ assumes trapped particles distribute uniformly along the poloidal direction and might amplify the drive of deeply-trapped fraction. In addition, the omitted barely-trapped EE-drive is in the higher order due to its small population. 

\subsection{Linear properties of beta-induced Alfven eigenmode driven by energetic electrons (e-BAE)}\label{4.2}

To verify the global fluid-kinetic hybrid simulation model and corresponding numerical implementation in MAS code, we carry out simulations of e-BAE in analytic geometry based on a well-established benchmark case \cite{cheng2016}, where the safety factor profile and concentric-circular geometry are shown in figures \ref{q} (a) and \ref{ana_trap} respectively. Protons are used for thermal ions with $Z_i=e$. The thermal electron, thermal ion and EE temperatures are uniform with $T_{i0} = T_{e0} = 500eV$ and $T_{h0} = 25 T_{e0}$, and the thermal electron density is uniform with $n_{e0} =1.3\times 10^{14}cm^{-3}$. The EE density profile is described by $n_{h0} = 0.05n_{e0}\left[1+0.2\left(tanh\left(0.26-\hat{\psi}\right)/0.06\right) -1.0\right]$, where $\hat{\psi}= \psi/\psi_w$ is the poloidal magnetic flux normalized by the wall value, and the reciprocal of EE density scale length $L_{n,h}^{-1} = \left(\nabla n_{h0}/n_{h0}\right)^{-1}$ peaks at $q=2$ rational surface with $R_0/L_{nh} = 12.7$. The thermal ion density is then determined by the quasi-neutrality condition $Z_in_{i0} + q_e\left(n_{e0} + n_{h0}\right)=0$. The on-axis magnetic field strength is $B_0 = 1.91T$, the major radius is $R_0 = 0.65m$, the minor radius is $a = 0.333R_0$, and the safety factor profile in figure \ref{q} (a) is $q = 1.797+0.8\hat{\psi}- 0.2\hat{\psi}^2$. 

\begin{figure}[H]
	\center
	\includegraphics[width=0.88\textwidth]{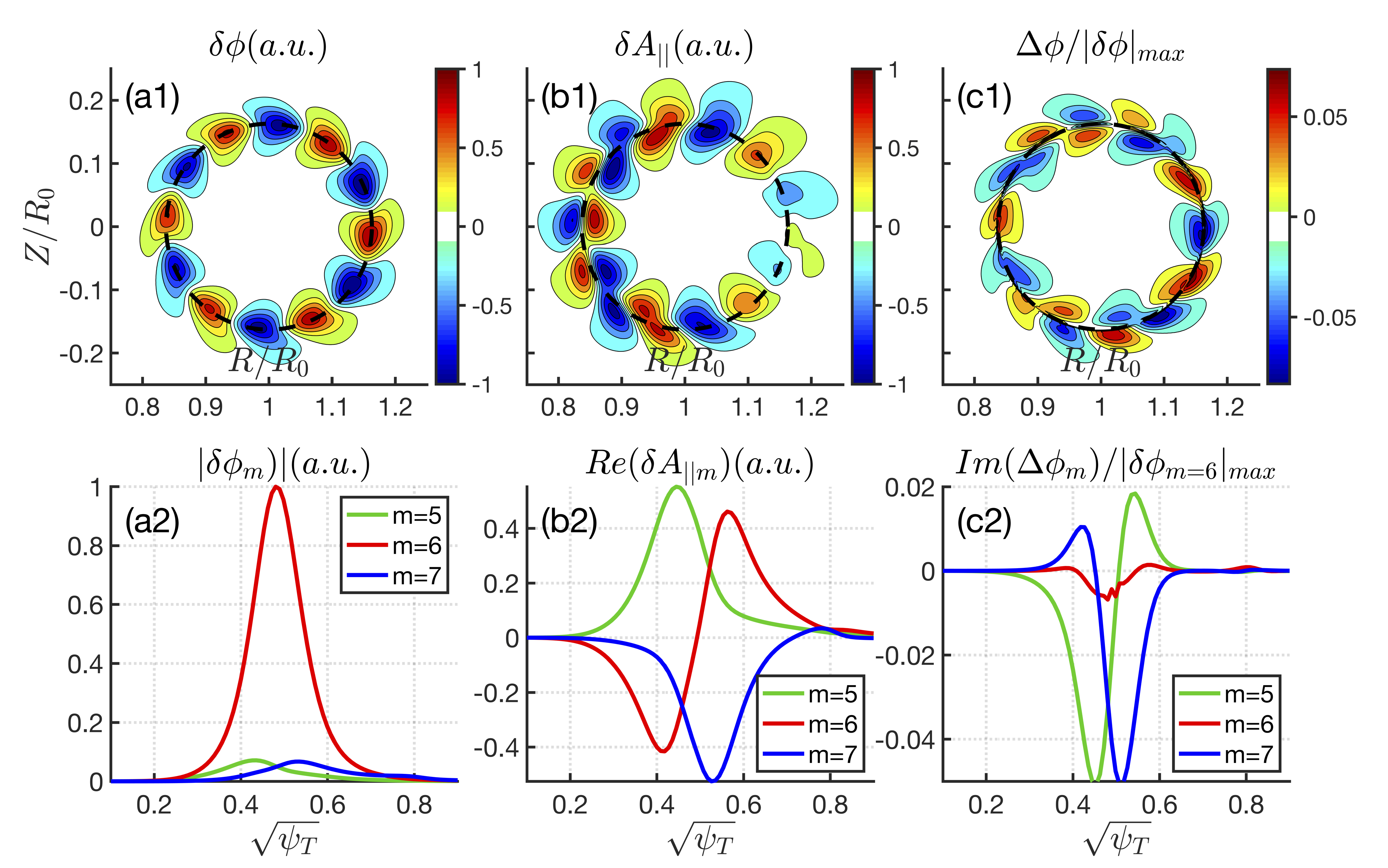}
	\caption{The 2D poloidal mode structures of (a1) electrostatic potential $\delta\phi$, (b1) parallel vector potential $\delta A_{||}$ and (c1) $\Delta\phi = \delta\phi-\delta\psi$ for $n=3$ e-BAE with the on-axis values of $n_{h0} = 0.05n_{e0}$ and $T_{h0} = 25T_{e0}$. (a2), (b2) and (c2) are the corresponding radial profiles of each poloidal harmonic.}
	\label{mode_struc}	
\end{figure}

\begin{figure}[H]
	\center
	\includegraphics[width=0.88\textwidth]{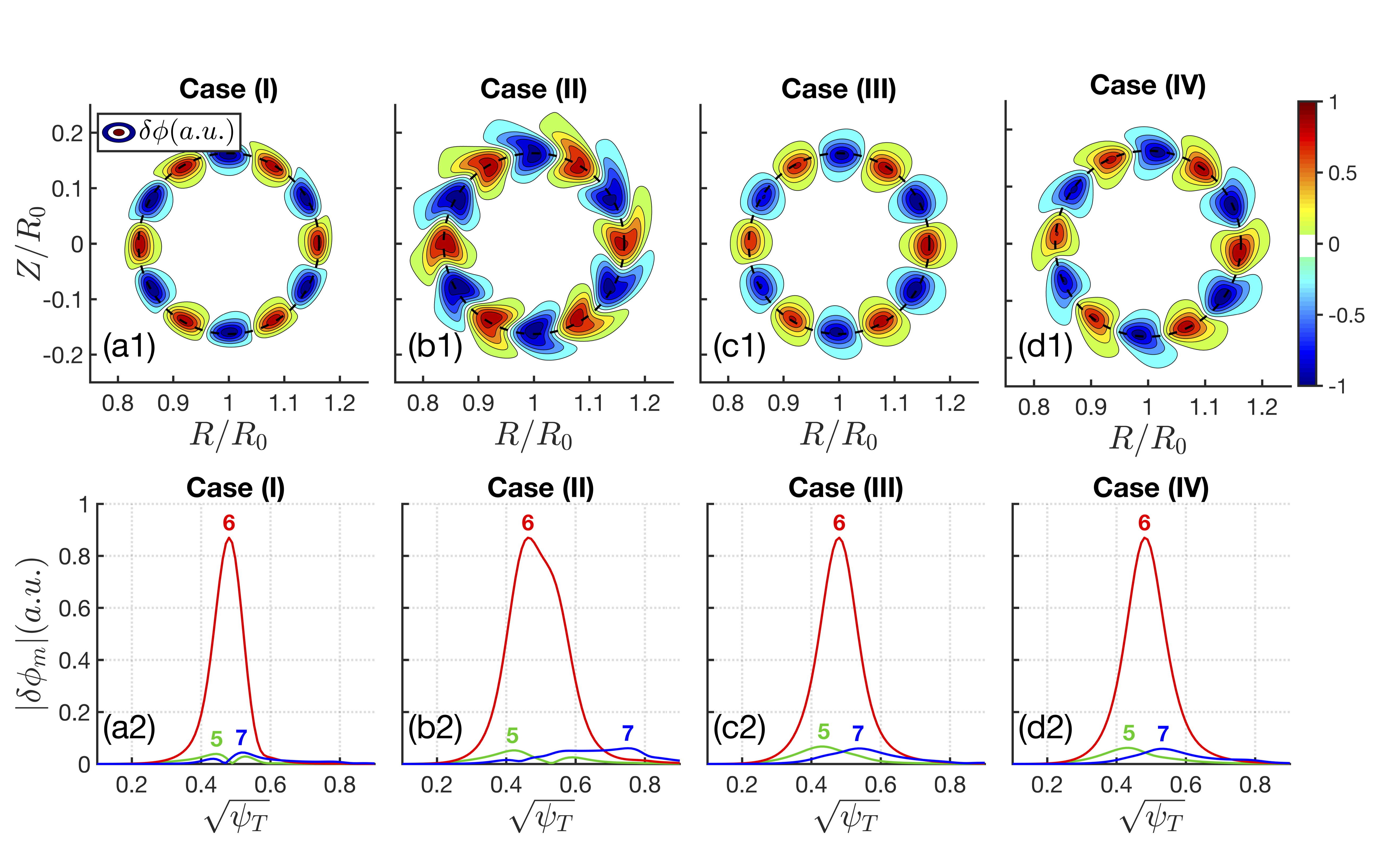}
	\caption{The 2D poloidal mode structures of electrostatic potential $\delta\phi$ with different EE terms in Eq. \eqref{vor_eq}. (a) Case (I): drop both EE-IC and EE-KPC terms. (b) Case (II): drop EE-IC term and keep EE-KPC term. (c) Case (III): keep EE-IC term and drop EE-KPC term. (d) Case (IV): keep both EE-IC and EE-KPC terms.}
	\label{mode4panel}	
\end{figure}

The global mode structure and dispersion relation of $n = 3$ e-BAE are analyzed as follows. The 2D poloidal mode structure of electrostatic potential $\delta\phi$ is shown in figure \ref{mode_struc} (a1), which is characterized with a ‘boomerang’ shape. The dominant principal poloidal harmonic of $m=6$ peaks at $q = 2$ rational surface, of which amplitude is much larger than the neighboring sideband harmonics of $m =5$ and $m=7$ that corresponds to a weakly ballooning structure as shown in figure \ref{mode_struc} (a2). Different from $\delta \phi$, the poloidal mode structure of parallel vector potential $\delta A_{||}$ exhibits anti-ballooning feature as shown in figure \ref{mode_struc} (b1), where the real parts of $m=5$ and $m=7$ poloidal sidebands are comparable to the resonant $m=6$ harmonic but with phase difference as shown in figure \ref{mode_struc} (b2). Moreover, the mode structure of $\Delta\phi = \delta\phi - \delta\psi$ is shown in figures \ref{mode_struc} (c1) and (c2), which represents the derivation from the ideal-MHD limit and reflects the polarization of each poloidal harmonics. The $m=6$ harmonic is predominantly Alfvenic according to the small $\Delta\phi$ since the electrostatic component $\delta\phi$ and electromagnetic component $\delta\psi$ nearly cancel out, while $\Delta\phi$ perturbation is large in $m=5$ and $m=7$ sidebands which indicates the acoustic component becomes more important. 
\begin{table}[H]
	\caption {Four simulation cases of $n=3$ e-BAE with different EE physics. Note that EE-IC and EE-KPC terms are given in Eq. \eqref{vor_eq}, which represent the fluid and kinetic EE responses respectively.} \label{four_cases} 
	\begin{center}
		\begin{tabular}{ l  l  l  l  l }
			\hline
			\rule{0pt}{15pt}
			Case & EE-IC & EE-KPC& $\omega_r\left(V_{Ap}/R_0\right)$& $\gamma\left(V_{Ap}/R_0\right)$\\ 
			\hline
			\rule{0pt}{15pt}
			(I) & No & No & 0.160 & -0.00707\\
			\rule{0pt}{15pt}
			(II) & No & Yes & 0.175 &0.00496   \\
			\rule{0pt}{15pt}
			(III) &  Yes & No & 0.134 & -0.00909  \\
			\rule{0pt}{15pt}
			(IV) & Yes & Yes & 0.149 & 0.00904   \\
			\hline
		\end{tabular}
	\end{center}
\end{table}
The symmetry breaking of AE mode structure due to non-ideal MHD effects, such as kinetic EPs induced anti-Hermitian part of dielectric tensor \cite{ma2015}, have been widely observed in simulations \cite{wang2013, lu2018} and experiments \cite{tobias2011, heidbrink2022}, which are characterized with twisted mode structures . To understand the role of EEs on ‘boomerang’ shape mode structure in figure \ref{mode_struc} (a) that is no longer up-down symmetric, four simulation cases are performed with different EE physics as described in table \ref{four_cases}. Case \{I\} corresponds to the Landau-fluid simulation of bulk plasma without any EE effects, and the poloidal mode structure is relatively up-down symmetric as shown in figure \ref{mode4panel} (a1), where the small distortion is due to the kinetic effects of bulk plasmas that lead to anti-Hermitian contribution. In case \{II\}, the EE-KPC term is added on top of case \{I\}, and the 2D mode structure is radially broadened and drastically twisted with obvious tails on both sides of mode rational surface in figure \ref{mode4panel} (b1), which indicates that the anti-Hermitian contribution from EE-KPC is much larger than bulk plasmas. For comparison, figure \ref{mode4panel} (c1) shows the 2D mode structure of case \{III\} that includes EE-IC term on top of case \{I\} instead of EE-KPC term, and it can be seen that EE-IC term has little impact on the symmetry breaking which is different from case \{II\}, but EE-IC term can broaden the radial width to a certain extent. In case \{IV\}, the non-perturbative effects of both EE-IC and EE-KPC terms are considered (figure \ref{mode4panel} (d1) and figure \ref{mode_struc} (a)), it is seen that the degree of symmetry breaking in case \{IV\} is between case \{II\} and case \{III\}, and the radial width is close to both case \{II\} and case \{III\}. Thus, there are two EE non-perturbative effects on e-BAE mode structure: (i) the EE-KPC term provides the dominant kinetic effects and induces the large ant-Hermitian part of dispersion relation, which is responsible for the distortion e-BAE mode structure; (ii) the EE-IC term represents the fluid-like convective response and enhances the Hermitian part, which not only compensates the symmetry breaking induced by EE-KPC term but also broadens the radial width through MHD interchange effect. 

\begin{figure}[H]
	\center
	\includegraphics[width=0.8\textwidth]{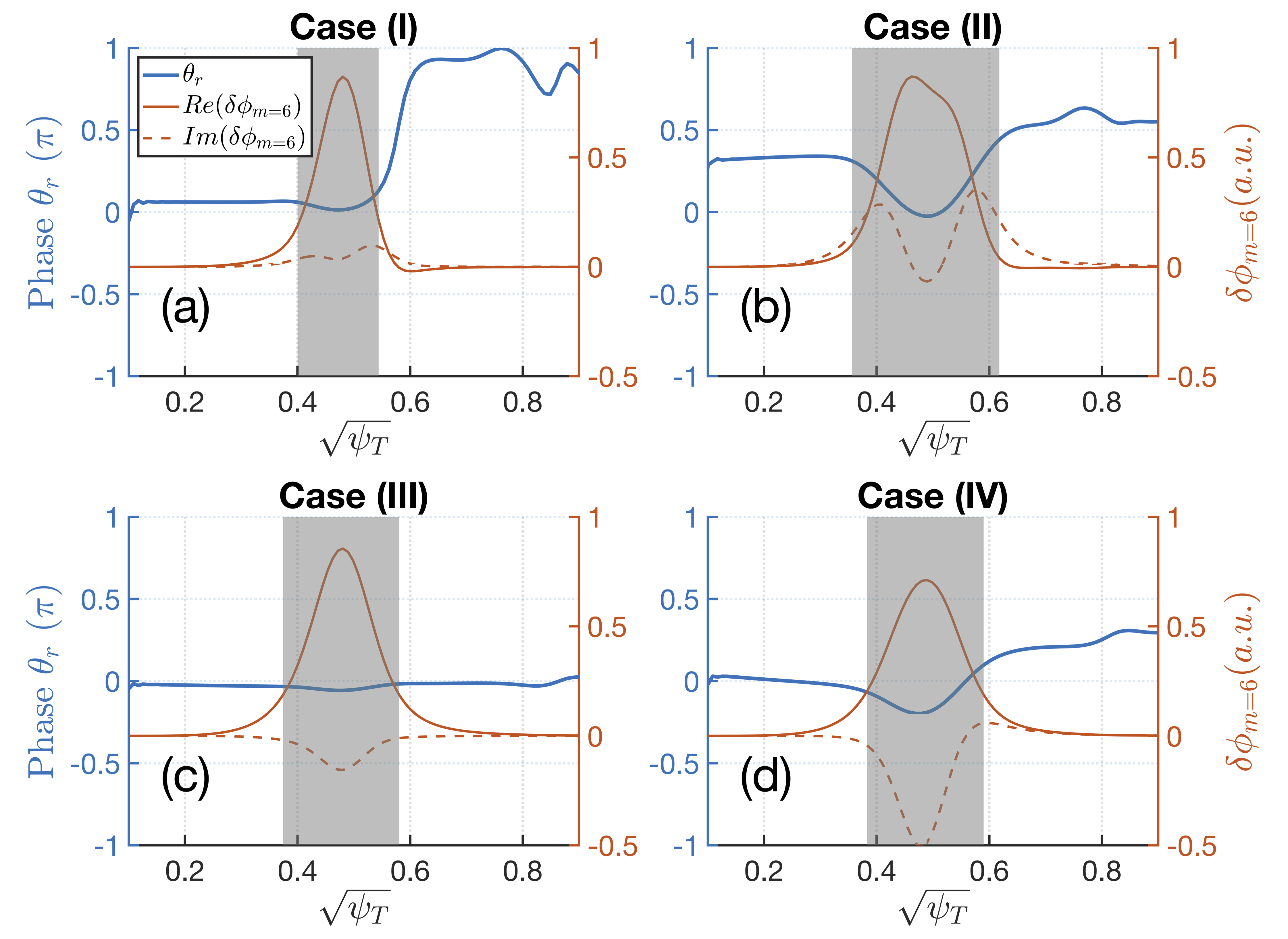}
	\caption{(a)-(d) The dominant $m=6$ harmonic radial profiles of electrostatic potential $\delta\phi$ and corresponding phase angle $\theta_r$ for cases (I)-(IV). The radial profile of $\theta_r = arctan(Im(\delta\phi_{m=6})/Re(\delta\phi_{m=6}))$ is indicated by the blue solid line, and the radial profiles of $Re\left(\delta\phi_{m=6}\right)$ and $Im\left(\delta\phi_{m=6}\right)$ are indicated by the red solid line and red dashed line respectively. The gray shaded region indicates the radial domain of $\delta\phi_{m=6}$ with high intensity.}
	\label{phase}	
\end{figure}

\begin{figure}[H]
	\center
	\includegraphics[width=0.5\textwidth]{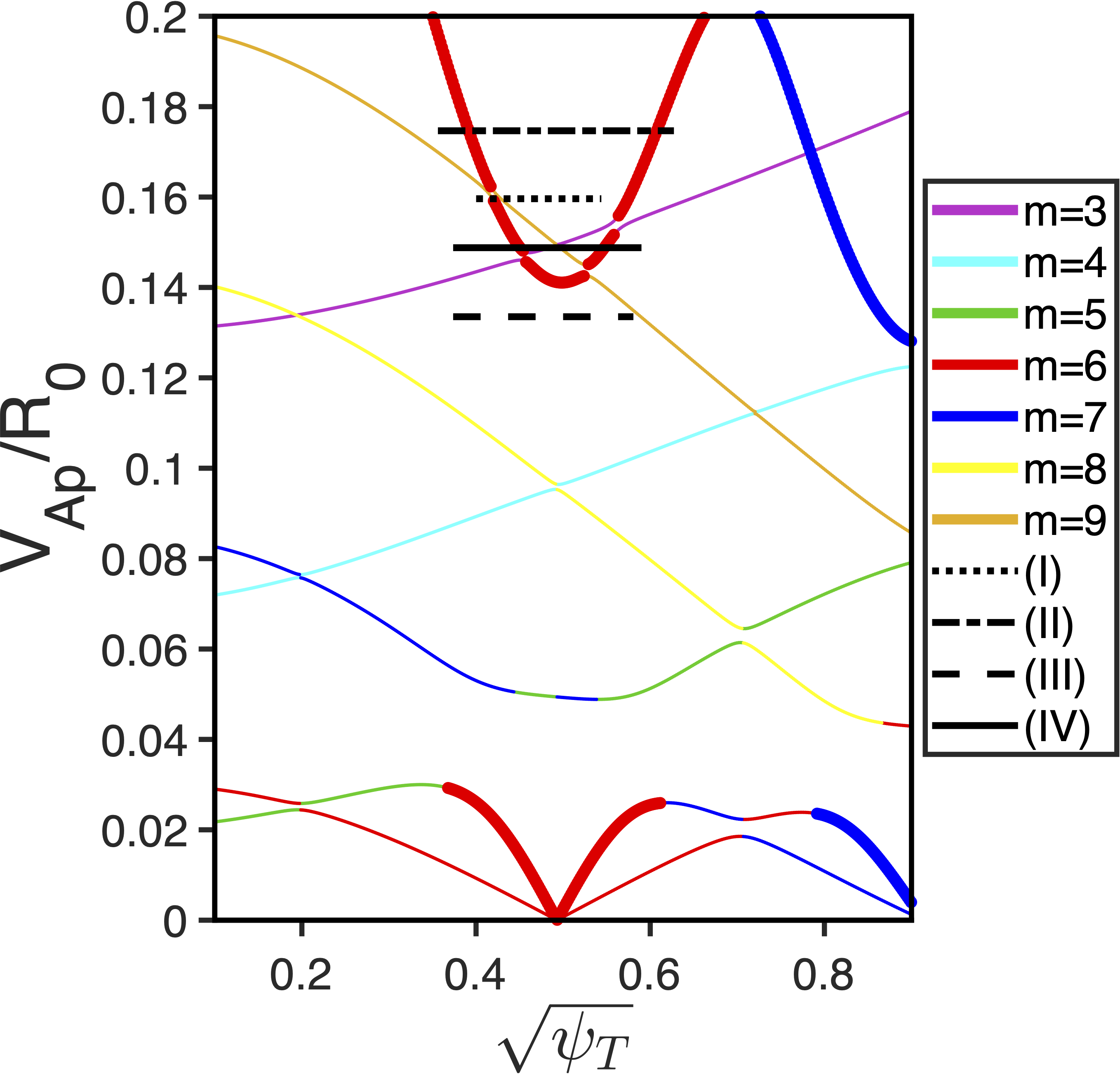}
	\caption{The e-BAE frequencies of cases (I)-(IV) and continuous spectra. The thick lines represent the Alfv\'enic branch and the thin lines represent the acoustic branch.}
	\label{continuum}	
\end{figure}

\begin{figure}[H]
	\center
	\includegraphics[width=1\textwidth]{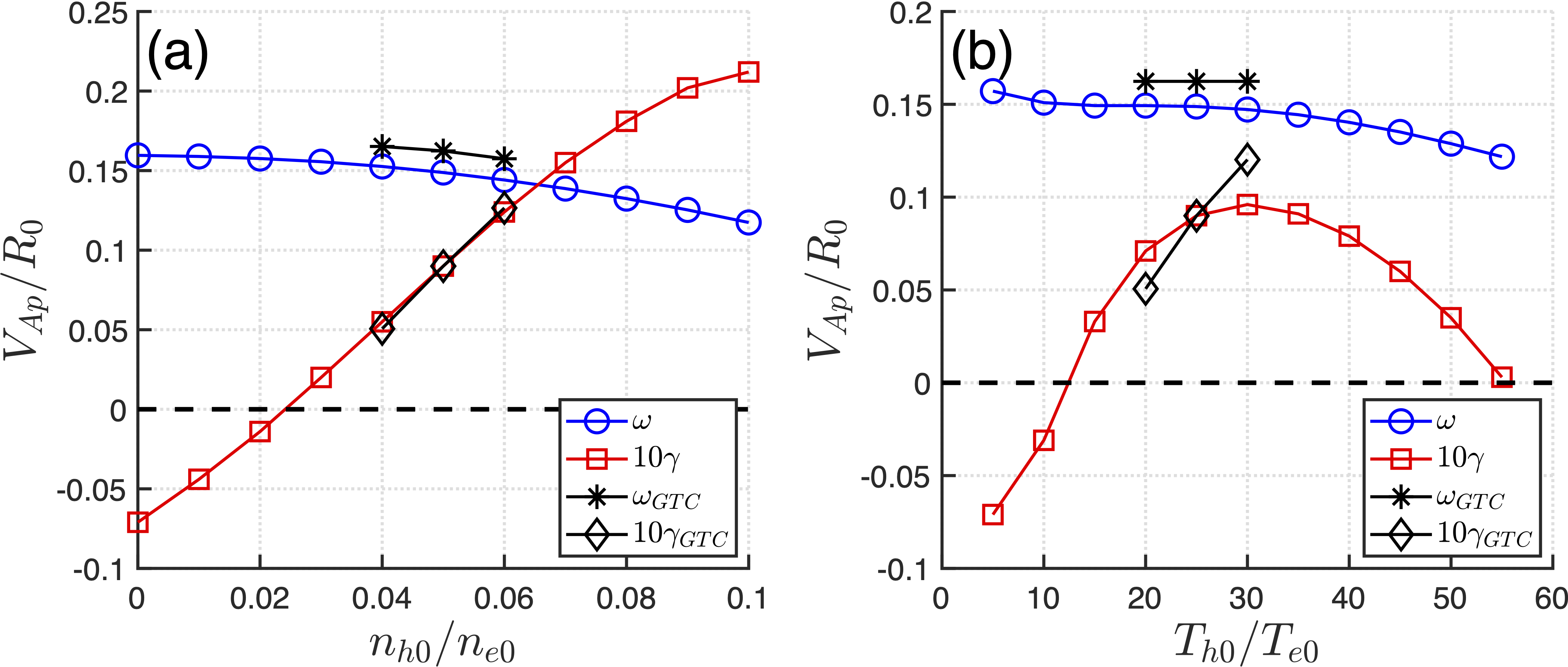}
	\caption{The comparisons of e-BAE frequency and growth rate between MAS (blue and red colors) and GTC (black color) simulations. (a) Scan the EE density $n_{h0}$ with fixed $T_{h0} = 25T_{e0}$. (b) Scan the EE temperature $T_{h0}$ with fixed $n_{h0} = 0.05n_{e0}$.}
	\label{DR}	
\end{figure}

On the other hand, the radial profile of phase angle $\theta_r$ can also reflect the mode structure symmetry, which has received interest from recent experiments \cite{tobias2011,heidbrink2022}. Here we combine $\theta_r$ radial profiles of cases \{I\}-\{IV\} with corresponding 2D mode structures to illustrate their underlying connections. Since the poloidal harmonics of e-BAE are weakly coupled, we choose the dominant $m=6$ harmonic of $\delta \phi$ and calculate $\theta_r$ according to $\delta\phi_{m=6} = |\delta\phi_{m=6}|exp(i\theta_r)$. In figure \ref{phase}, the blue solid line represents $\theta_r$, the red solid and red dashed lines represent the real and imaginary parts of $\delta\phi_{m=6}$ respectively, and the radial domain of e-BAE eigenfunction with finite amplitude is marked as a gray shaded region. As shown in figures \ref{phase} (a) and (c), $\theta_r$ profiles almost remain constant in the e-BAE region of cases \{I\} and \{III\}, which correspond to the relatively symmetric mode structures in figures \ref{mode4panel} (a1) and (c1) respectively. In contrast, $\theta_r$ profiles show large variations in the e-BAE region of cases \{II\} and \{IV\} due to the anti-Hermitian EE-KPC term that breaks the poloidal up-down symmetry. However, the poloidal phase shifts at different radial locations are approximately symmetric with respect to the e-BAE peak as shown in figures \ref{phase} (b) and (d) (because the EE gradient profile peaks at the mode rational surface in our simulations), which still indicate the radial symmetry of ‘boomerang’ shape mode structures characterized with two symmetric tails as shown in figures \ref{mode4panel} (b1) and (d1). In addition, we show the real frequencies, radial positions and mode widths of cases \{I\}-\{IV\} on corresponding $n=3$ continuous spectra in figure \ref{continuum}.
To further investigate e-BAE frequency and growth rate dependencies on EE-drive, we vary $n_{h0}$ and $T_{h0}$ in amplitudes while remaining profiles unchanged. The e-BAE frequency exhibits weak dependence on $n_{h0}$ and $T_{h0}$, which slowly decreases as $n_{h0}$ and $T_{h0}$ increase in figure \ref{DR}. In contrast, the growth rate is sensitive to both $n_{h0}$ and $T_{h0}$, which linearly increases with $n_{h0}$ due to the enhanced drive in figure \ref{DR} (a), and initially increases and then decreases with $T_{h0}$ in figure \ref{DR} (b) according to the requirement of resonance condition. The intersections of the black dashed lines and growth rate curves in figure \ref{DR} represent the marginal stable e-BAEs, of which $n_{h0}$ and $T_{h0}$ values give the excitation thresholds that EE-drive just overcomes the continuum damping \cite{berk1991,rosenbluth1992} and radiative damping/Landau damping of bulk plasma \cite{mett1992,fu2005,lauber2005}.
From figures \ref{mode_struc} and \ref{DR}, it is seen that both the mode structure and dispersion relation of e-BAE from MAS simulations quantitatively agree with the first-principle GTC results in figures 2 and 3 of Ref. \cite{cheng2016}, where the deeply-trapped EEs play a dominant role for e-BAE excitation.

\begin{figure}[H]
	\center
	\includegraphics[width=1\textwidth]{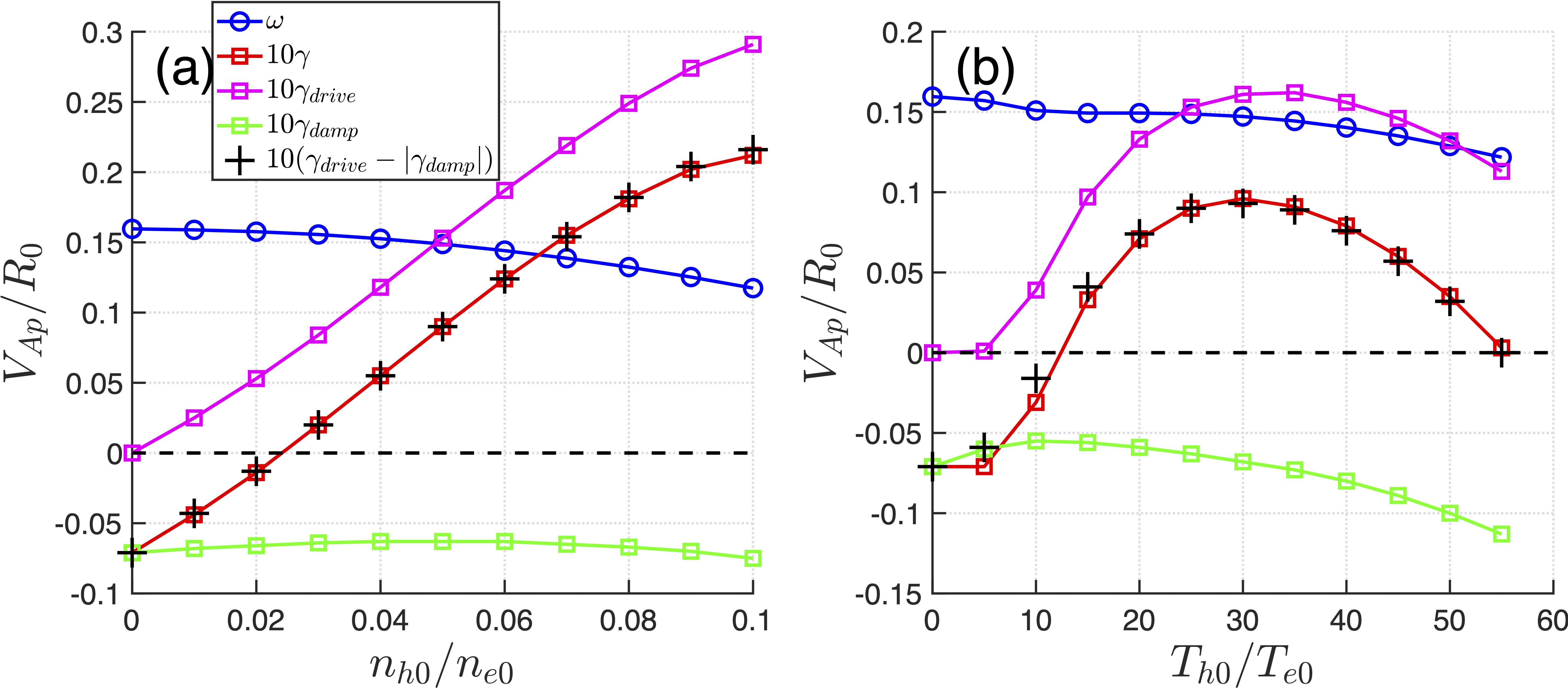}
	\caption{The kinetic EE drive $\gamma_{drive}$ (the magenta squares) and bulk plasma damping $\gamma_{damp}$ (the green squares) calculated by MAS. (a) Scan the EE density $n_{h0}$ with fixed $T_{h0} = 25T_{e0}$. (b) Scan the EE temperature $T_{h0}$ with fixed $n_{h0} = 0.05n_{e0}$. The individual calculations of $\gamma_{drive}$ and $\gamma_{damp}$ are consistent with the gross growth rate $\gamma$ (the red squares) obtained from simulations that incorporate both EE drive and bulk plasma damping.}
	\label{DR_damp_drive}	
\end{figure}

Moreover, in order to clarify the competition between EE excitation and bulk plasma damping, we compute the EE drive $\gamma_{drive}$ in the absence of dissipation by droping Landau damping terms associated to $|k_{||}|$ in Eqs. \eqref{thermal_e_ohm}, \eqref{ion_pressure} and \eqref{uipara2}, and compute the bulk plasma damping rate $\gamma_{damp}$ by isolating EE drive effect (i.e., only keeping the real part of EE response functions in Eqs. \eqref{nee_NA} and \eqref{Pee_NA} while droping the imaginary part), and $\gamma_{drive}$ and $\gamma_{damp}$ dependencies on $n_{h0}$ and $T_{h0}$ are shown in figure \ref{DR_damp_drive}. It should be noted that both bulk plasma and EE non-perturbative effects, namely, modifications on mode structure and real frequency, are kept for calculating $\gamma_{drive}$ and/or $\gamma_{damp}$. For example, $\gamma_{damp}$ shows relatively strong dependency on $T_{h0}$ in figure \ref{DR_damp_drive} (b), which indicates that EE non-perturbative effect can quantitatively affect the bulk plasma damping on e-BAE through altering the mode structure and real frequency, and thus becomes necessary for accurate damping and growth rate calculation. Meanwhile, the $\gamma_{drive} - |\gamma_{damp}|$ agrees well with the gross growth rate $\gamma$ using comprehensive model with both EE drive and bulk plasma damping, which confirms the correctness of our method for obtaining $\gamma_{drive}$ and $\gamma_{damp}$ that covers not only the unstable regime of $|\gamma_{damp}|\ll\gamma_{drive}$ but also the marginally stable regime of $|\gamma_{damp}|\sim\gamma_{drive}$, and thus provides useful damping and drive information for linear stability and further nonlinear dynamics analyses \cite{berk1996}.


\section{Conclusions}\label{5}

In summary, we have formulated a novel fluid-kinetic hybrid model that couples drift-kinetic EEs to the Landau-fluid bulk plasmas in general geometry, which retains the deeply-trapped EE physics in the simulations of kinetic-MHD processes in a non-perturbative manner. The new model has been implemented and verified in the eigenvalue code MAS \cite{bao2022} with practical applications that cover MHD modes, AEs and drift wave instabilities, and the main characteristics are summarized as follows.
\begin{enumerate}
	
	\item \textbf{Drift-kinetic description of EE responses.} The EE perturbed distribution is solved from the drift-kinetic equation with well-circulating approximation for passing EEs and deeply-trapped approximation for trapped EEs, which keeps the EE kinetic effect of precession drift resonance that is responsible for the excitations of most EE-driven AEs, and the EE fluid effects such as adiabatic and convective responses. Although the well-circulating and deeply-trapped approximations are made in the derivation, it is shown that the EE moments integrated from the perturbed distribution, i.e., perturbed density, parallel current and pressure, can well guarantee the conservation property of EE continuity equation.

	\item \textbf{Improved deeply-trapped model.} The deeply-trapped approximation is applied for deriving the EE-drive terms with dominant precessional drift resonance, which has computational advantage because the heavily numerical integration along the realistic particle orbit is not involved. Specifically, this approximation is made for both the calculation of precession frequency (i.e., $\overline{\omega}_d$) and bounce-average operations on electromagnetic fields (i.e., $\overline{\delta\phi}$, $\overline{\delta\psi}$ and $\overline{\omega_{d}\delta\psi}$), where we induce a control parameter $\lambda_{low}$ to calculate the 2D poloidal profile of deeply-trapped particle fraction $f_t$, namely, only keep the trapped particles that satisfy $\overline{\omega}_d/\omega_{d0}\sim 1$ and $|\theta|\ll1/|nq-m|$ simultaneously. By improving the accuracy of deeply-trapped approximation with above two constraints, MAS simulations of e-BAE mode structure and dispersion relation show good agreements with gyrokinetic PIC simulations \cite{cheng2016}.
	
	\item \textbf{Non-perturbative approach.} The EEs are self-consistently incorporated in MAS computations of mode structure, real frequency and growth rate. The non-perturbative effects of EEs on e-BAE mode structures and corresponding radial variations of phase angle profiles are demonstrated. In particular, the EE-KPC term is found to effectively twist the e-BAE global mode structure and break the poloidal up-down symmetry due to the anti-Hermitian contributions to the dielectric tensor, while EE-IC term represents the fluid-like convective response and thus compensate the Hermitian part, which leads to the mode structure to be more symmetric. The EE-KPC term is also responsible for the large radial variations of phase angle, of which poloidal phase shifts at different radial locations are consistent with the ‘boomerang’ shape mode structure.
	
	\item \textbf{Comprehensive damping and driving effects.} The original Landau-fluid model for bulk plasmas in MAS \cite{bao2022} has already incorporated the important continuum damping, Landau damping and radiative damping. The newly formulated fluid-kinetic hybrid model combines the EE-drive together with various damping mechanisms, which is more reliable for explaining the marginal stable AEs observed in experiments and is useful for evaluating the EE density and temperature thresholds for AE excitation. For comparison, the bulk plasma dampings are commonly absent in traditional kinetic-MHD hybrid simulations that might affect the unstable spectra. 
	
\end{enumerate}

With the high efficiency of eigenvalue approach in computation, the upgraded MAS is suitable for fast parameter scans of deeply-trapped EE relevant problems that attract attentions in experiments. Meanwhile, the coupling scheme of EE and bulk plasma can be used for other distributions and/or numerical integration along realistic orbits in phase space, e.g., showing-down distribution and barely-trapped EEs.



\section{Acknowledgments}
This work is supported by the National MCF Energy R\&D Program under Grant Nos. 2018YFE0304100; National Natural Science Foundation of China under Grant Nos. 12275351, 11905290 and 11835016; and the start-up funding of Institute of Physics, Chinese Academy of Sciences under Grant No. Y9K5011R21. We would like to thank useful discussions with Zhixin Lu, Huishan Cai, Yong Xiao and Ruirui Ma.

\appendix

\end{document}